\documentclass[12pt,preprint]{aastex} 

\IfFileExists{srcltx.sty}{\usepackage[active]{srcltx}} 

\slugcomment{Submitted to the {\it Astrophysical Journal}} 
\shorttitle{NGC 4472 Abundance Pattern}
\shortauthors{Loewenstein and Davis}

\begin{document}
\title{The Abundance Pattern in the Hot ISM of NGC 4472: Insights and
Anomalies}
\author{Michael Loewenstein\altaffilmark{1}}
\affil{Department of Astronomy, University of Maryland, College Park, MD 20742}
\email{Michael.Loewenstein.1@nasa.gov}

 \and

\author{David S. Davis\altaffilmark{1,2}} 
\affil{Department of Physics, University of Maryland Baltimore County,
Baltimore, MD 21250}
\email{David.S.Davis@nasa.gov}

\altaffiltext{1}{CRESST and X-ray Astrophysics Laboratory, NASA/GSFC,
Greenbelt, MD.}
\altaffiltext{2}{CRESST and the Astroparticle Physics Laboratory,
NASA/GSFC, Greenbelt, MD.}


\begin{abstract}
Important clues to the chemical and dynamical history of elliptical
galaxies are encoded in the abundances of heavy elements in the X-ray
emitting plasma. We derive the hot ISM abundance pattern in inner
($0-2.3R_e$) and outer ($2.3-4.6R_e$) regions of NGC 4472 from
analysis of {\it Suzaku} spectra, supported by analysis of co-spatial
{\it XMM-Newton} spectra. The low background and relatively sharp
spectral resolution of the {\it Suzaku} XIS detectors, combined with
the high luminosity and temperature in NGC 4472, enable us to derive a
particularly extensive abundance pattern that encompasses O, Ne, Mg,
Al, Si, S, Ar, Ca, Fe, and Ni in both regions. We apply simple
chemical evolution models to these data, and conclude that the
abundances are best explained by a combination of $\alpha$-element
enhanced stellar mass loss and direct injection of Type Ia supernova
(SNIa) ejecta. We thus confirm the inference, based on optical data,
that the stars in elliptical galaxies have supersolar $[\alpha/Fe]$
ratios, but find that that the present-day SNIa rate is $\sim 4-6$
times lower than the standard value. We find SNIa yield sets that
reproduce Ca and Ar, or Ni, but not all three simultaneously. The low
abundance of O relative to Ne and Mg implies that standard core
collapse nucleosynthesis models overproduce O by $\sim 2$.
\end{abstract}



\section{Introduction}

\subsection{Context}

Galaxy formation encompasses the processes of star formation and
largescale dynamics -- where dynamics is defined in its broadest sense
to include the assembly of dark matter and baryons, as well as the
exchange of mass and energy among the various galactic components and
with the external environment. Star formation ultimately leads to the
nuclear production of metals, while dynamical processes determine
their destination. Therefore the abundance, abundance pattern, and
location of metals in different galaxies in the {\it local} universe
provide fundamental insights into the construction and development of
galaxies..

Chemical evolution modeling quantifies this connection. Heavy elements
are both statically and explosively synthesized in evolving stars
following episodes of star formation with abundances that reflects the
total mass of stars formed and the initial mass function (IMF), as
well as the production rate of Type Ia supernova (SNIa) progenitor
binary stars. These metals are subsequently returned to the
interstellar medium (ISM), and the enriched ISM may regenerate stars
or escape into intergalactic space, depending on the level of
supernova energy injection that accompanies the metal enrichment --
and how efficiently that energy is channeled into outflow. These
processes must depend on initial conditions and environment in such a
way as to produce the diverse and evolving universe of galaxies that
we observe from the Local Group out to the distant universe at
redshifts of 6 and greater. One can illuminate the history in the
ensemble of galaxies in a cluster by measuring abundances in the
intracluster medium (ICM; see, e.g., Loewenstein 2006), and in
individual galaxies by measuring stellar and interstellar abundances.

It is evident that in giant elliptical galaxies, predominantly
composed of stellar populations passively evolving since the early
universe, the star-gas cycle was greatly accelerated and concentrated
in time relative to the Milky Way and other late-type galaxies
\citep{pap06,vv07,jim07,per08}. Moreover, the prodigious metal content
in the intergalactic medium of galaxy clusters dominated by
ellipticals implies that strong galactic winds were driven in these
galaxies. However, details such as the precise initial epoch and
duration of star formation, the form of the IMF, the relative roles of
SNIa and core collapse (SNII) supernovae in enriching and expelling
ISM -- and how all these depend on mass and environment -- remain
unclear.

Abundances in the stellar component and their correlation with
structural parameters, provide key diagnostics of elliptical galaxy
ages, star formation and accretion histories, and IMFs
\citep{tho05,pip09,cle09,cpm09,ts09,rec09}. Since stars dominate the
baryon content of ellipticals, the level of stellar $\alpha$-element
enrichment is determined by the past integrated frequency of SNII, and
hence the number of massive stars formed. For a given total mass in
stars, this then constrains the IMF. The stellar $\alpha/Fe$ ratio is
then determined by the relative contributions of SNIa and SNII.
Careful modeling of multiband photometry and very high
signal-to-noise optical spectroscopy can, in principle, determine the
abundances as well as the ages of ellipticals. However, since
interpretation of colors and absorption line indices suffer from a
degeneracy in their dependence on the SFH and overall metallicity --
even for an assumed IMF and distribution of abundance ratios
\citep{how05} -- optical studies of the composite stellar population
cannot clearly and unambiguously measure these fundamental quantities.
The additional consideration of Balmer emission line indices can
assist in breaking some of the degeneracies \citep{t00} but, e.g.,
does not easily distinguish a very old stellar population with a
``frosting'' of recent star formation episodes from a somewhat younger
one \citep{st07,ts09}. Since, optical line indices can only be
interpreted within the context of a conjectured SFH and IMF, they are
more robust in providing consistency tests of combinations of ages and
abundance ratios ({\it i.e.}, Mg/Fe) than they are in determining
absolute abundances \citep{how05}. Moreover, the effects of variations
in individual elements is not easily disentangled \citep{lee09}, and
optical absorption features are only measured out to 1-2 (projected)
effective radii ($R_e$) -- the abundances in, typically, half of the
stellar mass are unknown.

Fortunately, mass lost by post-main-sequence stars in elliptical
galaxies is heated to millions of degrees K as it is incorporated into
the interstellar medium (ISM), where it is amenable to relatively
robust and straightforward X-ray spectroscopic analysis
\citep{mb03}. Because the gas, with the possible exception of a few
high oscillator strength lines, is optically thin the absolute
abundances of a wide range of elements with prominent emission
features that can include C, N, O, Ne, Mg, Al, Si, S, Ar, Ca, Fe, and
Ni may be directly derived given sufficient spectral resolution,
sensitivity, and bandpass. Because this approach relies on strong
emission lines, X-ray derived abundances may be determined out to very
large radii -- often extending to the edge of the optical galaxy and
beyond \citep{mat98}.

While hot ISM abundances may be more directly derived than stellar
abundances, their interpretation requires careful deconstruction
within the context of physical gasdynamical and chemical evolutionary
models. Unlike the case of the ICM that dominates the baryon content
in clusters, most of the metals produced by the evolving stellar
population are expelled from galaxies or locked up in stars -- not
accumulated in a reservoir of hot gas. Enrichment timescales for
elements synthesized by SNIa, SNII, and intermediate mass stars are
distinct, and coupled in a complex way to star formation, mass return,
and outflow timescales. Nevertheless, one can employ reasonable
assumptions, approximations, and simplifications to construct simple
models that track the evolution of global abundances of these
elements. Comparison of these models with observations constrain
important features of the nature of the galaxy stellar population, as
well as the rate and ultimate disposition of metals ejected as this
population evolves. As such, they may serve as guides to subsequent
more complex modeling.

In this paper, we introduce such models and apply them to the hot ISM
abundances in NGC 4472 that we infer from analysis of {\it Suzaku} and
{\it XMM-Newton} spectra. Because of its brightness and temperature
structure, a particularly well-determined and wide-ranging X-ray
abundance pattern is measurable in NGC 4472 -- one that is made more
accessible utilizing the low internal background and sharp energy
resolution of the {\it Suzaku} XIS detectors.

\subsection{Brief Survey of Previous X-ray Results}

High quality X-ray spectra for a large sample of elliptical galaxies
was first obtained with the {\it Advanced Satellite for Cosmology and
Astrophysics} ({\it ASCA}). Fe abundances in most of the X-ray
brightest ellipticals were found to be within a factor of two of
solar, with no strong evidence of non-solar ratios of
$\alpha$-elements to Fe \citep{bf98,mom01}. X-ray spectra extracted
from the {\it Chandra} and {\it XMM-Newton} CCD detectors take
advantage of improved sensitivity and spectral resolution, and far
superior angular resolution, to more cleanly derive abundance patterns
and gradients (Humphrey \& Buote 2006, and references therein). {\it
Chandra} results indicate that ISM Fe abundances are roughly solar and
decline slowly with radius as far out as tens of kpc
\citep{hb06,ath07}. In addition, approximately solar Mg/Fe and Si/Fe
ratios and subsolar O/Fe ratios are inferred.

Abundances derived from the {\it XMM-Newton} reflection grating
spectrometer (RGS) benefit from cleaner separation of individual
features, so that effects of resonance scattering on optically thick
lines, and multiphase gas on temperature-sensitive lines and line
ratios, may be directly addressed. Analysis of {\it XMM-Newton}
reflection grating spectrometer (RGS) spectra of NGC 4636 by
\cite{xu02} yields subsolar Mg/Fe and O/Fe ratios, a roughly solar
Ne/Fe ratio, and a supersolar N/Fe ratio. The difficulty in analyzing
grating spectra of extended sources, and limitations in sensitivity
and bandpass, restrict the applicability of RGS spectroscopy to X-ray
luminous galaxies with the highest central X-ray surface
brightnesses. \cite{ji09} analyze {\it Chandra} and {\it XMM-Newton}
EPIC and RGS data in 10 bright systems. They confirm the above
abundance pattern with respect to O/Mg/Si/Fe, and extend to Ne and S
that show no strong deviations from solar ratios, and Ni that is
generally supersolar.

The lower internal background and sharper energy resolution of the
{\it Suzaku} XIS CCDs enable the derivation of abundance patterns to
larger radii, and allow for a more accurate measurement of possible
features at high energy originating from S, Ar, Ca, and Ni. The {\it
Suzaku} low energy sensitivity makes it suitable for measuring O as
well. Results on NGC 1399 \citep{mat07}, NGC 720 \citep{taw08}, NGC
5044 \citep{kom09}, NGC 507 \citep{sat09}, and NGC 4636 \citep{hay09}
confirm the solar Si/Fe and subsolar O/Fe ratios, demonstrate that
these persist to large radii, and extend this behavior to S/Fe. It is
the case that this is also broadly true of Mg/Fe; however, Mg/Fe is
somewhat lower in NGC 720 and NGC 1399. Ne/Fe seemingly shows more
variation, coming in at solar in NGC 4636, subsolar in NGC 720 and
supersolar in NGC 507 and NGC 5044. Ne does not seem to trace O as
might be expected.

Optical spectra imply that the stellar ratio of $\alpha$ elements to
Fe, expressed logarithmically with respect to solar as $[\alpha/Fe]$,
is such that $[\alpha/Fe]\sim 0.3-0.5$ -- although there is little
known about this ratio well beyond $R_e$ or about the relative
abundances of different $\alpha$ elements. If stellar mass loss
dominates ISM enrichment, they should display the same
$\alpha$-element overabundance. Based on the the above summary, this
is clearly not the case. Direct injection of SNIa will skew that
abundance pattern towards lower $[\alpha/Fe]$, but should do so in a
way that does not effect relative abundances of elements with low SNIa
nucleosynthetic yields such as O, Ne, and Mg. The ratios, with respect
to Fe, of Si, S, Ar, and Ca may be less affected, depending on the
level of SNIa enrichment since these elements are produced with
moderate efficiency by SNIa. The solar ratio seen for most of the
$\alpha$ elements in the hot ISM indicates a significant role for SNIa
enrichment; but, the fact that O is particularly underabundant while
Mg and Ne are as abundant as Si and S does not comfortably fit into
this scheme. We examine these issues more quantitatively below with
respect to the ISM abundances in NGC 4472, where measurements of Ca
and Ar in the ISM add a new twist to attempts to understand the
enrichment of this galaxy.

\section{Construction of NGC 4472 Spectra and Associated Files}

\subsection{{\it Suzaku} Spectral Extraction}

NGC 4472 was observed with {\it Suzaku} \citep{mit07} between
2006-12-03 and 2006-12-06 (OBSID=801064010) with an on-source exposure
time of 121 ksec -- 94 (27) ksec in $3\times 3$ ($5\times 5$) CCD
editing mode. At the time of observation, three co-aligned,
$17.8'\times 17.8'$ field-of-view X-ray Imaging Spectrometer (XIS) CCD
cameras \citep{koy07} -- two front-illuminated (FI: XIS0 and XIS3) and
one back-illuminated (BI:XIS1) -- were operational, each XIS in the
focal plane of an X-ray Telescope (XRT) with a $2'$ half-power
diameter \citep{ser07}.

Observations were conducted utilizing the space-row charge injection
(SCI) technique that reverses the degradation in energy resolution
caused by accumulated radiation damage \citep{nak08}. We initiate our
data reduction with the unfiltered event files. These data underwent
Version 2.0.6.13 pipeline processing on 2007-08-18 that enables one to
properly account for the effect of SCI on instrument characteristics
and performance \citep{uch08}. We reprocess the unfiltered event files
by hand in order to apply updated calibration data and software. Our
analysis generally follows the procedures outlined in ``The Suzaku
Data Reduction Guide,''
\footnote{http://heasarc.gsfc.nasa.gov/docs/suzaku/analysis/abc/} as
implemented in HEAsoft version
6.5.1.\footnote{http://heasarc.gsfc.nasa.gov/docs/software/lheasoft/}
We recalculate PI values and grades, select event grades (0, 2, 3, 4,
6) that correspond to X-ray photon events, filter on pixel status
(eliminating bad charge transfer efficiency columns, and rows
invalidated by the charge injection process) and select good time
intervals (GTI) based on pointing, data and telemetry rates, SAA
proximity (``$\rm{SAA}\_ {\rm HXD}\equiv 0, \rm{T}\_ \rm{SAA}\_
\rm{HXD}> 436$''), and proximity to the earth's limb and illuminated
Earth (``$\rm{ELV}> 5, \rm{DYE}\_ \rm{ELV}> 20$''). In addition,
telemetry-saturated frames and calibration source photons are screened
out; and, hot and flickering pixels are removed. Finally, we accept
only GTI where the revised geomagnetic cut-off rigidity COR2$>4$, thus
eliminating intervals with the highest particle background level
\citep{tawa08} without compromising overall statistical accuracy
(experiments with the stricter criteria $COR2>6$ yields identical
results). $5\times 5$ event files are converted to $3\times 3$ mode
format, and merged with the $3\times 3$ event files.

Spectra are extracted from $0-4'$ ($0-2.3R_e$) circular (inner) and
$4-8'$ ($2.3-4.6R_e$) annular (outer) regions, centered on the NGC
4472 optical nucleus ($\alpha=12~29~46.7$, $\delta=+08~00~02$) that
very closely corresponds to the X-ray peak in the {\it Suzaku}
image. Relatively large regions are chosen to minimize PSF effects,
and maximize the accuracy and range of abundance measurements. Based
on the simulations in \cite{sat07}, only a small fraction of counts in
these regions originate in other annuli in the Abell 1060 galaxy
cluster; and, this should be even more true for NGC 4472 based on its
relative surface brightness.

The spectral redistribution matrix files ({\bf rmf}) are generated
using {\bf xisrmfgen} version 2009-02-28. The {\bf rmf} and spectral
files are binned to 2048 channels. The effective area function files
({\bf arf}) for the source spectra are generated by the {\bf
 xissimarfgen} version 2009-01-08 Monte Carlo ray-tracing program
\citep{ish07} with 400000 simulation photons per energy bin, and an
input source fits image file generated from a $\beta$-model fit to the
background-subtracted X-ray surface brightness profile extracted from
archival {\it Chandra} data ($\beta=0.44$ and core radius $4''$,
although the results proved insensitive in experiments with larger
core radii ranging up to $\sim 1'$ and/or $\beta=0.48$). Spectra from
the FI chips, XIS0 and XIS3, are co-added and a weighted XIS0+3
response functions calculated from their respective {\bf rmf} and {\bf
 arf} files. Source spectra are grouped into bins with a minimum of
15 cts. Best-fit parameters and parameter uncertainties are derived
using $\chi^2$ statistics; experiments with the Cash statistic yielded
very similar results.

The XIS background includes contributions from Non-X-ray (charged
particle) Background (NXB), Galactic X-ray Background (GXB), and
(extragalactic) Cosmic X-ray Background (CXB). Since NGC 4472 fills
the {\it Suzaku} field of view, we estimate and subtract the NXB
component and include an additional thermal component in our spectral
fits to account for the GXB. The CXB has a similar shape, but a much
smaller magnitude in the inner region, than the intrinsic flux from
low-mass X-ray binaries (LMXBs) and is subsumed into that spectral
component. The NXB component is estimated from observations of the
night earth taken in SCI mode within 150 days of the starting or
ending dates of our observation using {\bf xisnxbgen} version
2008-03-08. The NXB event list in that time interval undergoes the
identical screening as the source data, is sorted by geomagnetic
cut-off rigidity, and weighted according to the cut-off rigidity
distribution in the source event file \citep{tawa08}. The estimated
NXB spectra include only those events collected in the regions on the
detector from which the source spectra are extracted. Since the NXB
has a spatial distribution distinct from that of the NGC 4472 X-ray
emission, a separate {\bf arf} file is generated using 2000000
simulation photons per energy bin from a uniform source of radius
$20'$ and applied to the background in spectral fits. Combined XIS0+3
NXB spectra and response functions are constructed as described above
for the source spectra.

The effective exposure time of the XIS0+3 (XIS1) spectrum is 232 ks
(117 ks). The total 0.3-12 keV band counts are 132032 (inner) and
70286 (outer) for XIS0+3, and 107232 (inner) and 72758 (outer) for
XIS1. The fraction of the total count rate in the NXB is 1.4\% (inner)
and 8.3\% (outer) for XIS0+3 , and 1.8\% (inner) and 10.3\% (outer)
for XIS1. Even in the outer region, its contribution is small in the
region where spectral features of interest lie (see Figure 3, below).

\subsection{{\it XMM-Newton} Spectral Extraction}

NGC~4472 was observed for $\sim$90 ksec with {\it XMM-Newton}. After
cleaning the dataset (the last 20 ksec of the full-field light curve
showed significant flares), $\sim$81 ksec ($\sim$72 ksec) of useful
observation time remained for the EPIC MOS1 and MOS2 (pn)
detectors. The data were reduced and extracted using SAS 6.5.0. The
latest calibration products were used to determine the response files
for the spectral\ analysis. The extracted spectra are binned so that
each channel has a minimum of 20 counts, and we consider energy ranges
of 0.4--7.0 keV and 0.32--2.0 keV for the EPIC cameras and RGS
detectors, respectively. To account for the fact that NGC~4472 is an
extended source, we used the rgsxsrc model in {\sc Xspec 11},
extracting the {\it Chandra} ACIS-S image to a create a surface
brightness distribution model to convolve with the RGS response. The
background for the central region was taken from the blank sky
background files provided by the {\it XMM-Newton} Science
Center. These were scaled to match the EPIC and RGS high energy
continua. EPIC MOS1, MOS2, and pn spectra are extracted from the inner
and outer regions described above. Subdividing the inner region
($r<200{^\prime}^\prime$) into multiple annular regions, and fitting
the corresponding EPIC spectra, reveals no significant $\sim$arcminute
gradients in abundance ratios (Section 3.2)

\section{Spectral Analysis}

\subsection{\it Suzaku Spectral Analysis}

The NGC 4472 {\it Suzaku} XIS0+3 and XIS1 spectra are simultaneously
fit in the 0.45-7.0 keV band using {\sc Xspec 12.5} with a source
model consisting of a two-temperature thermal plasma ({\bf vapec})
model with a single set of heavy element abundances (C, N, O, Ne, Mg,
Al, Si, S, Ar, Ca, Fe, Ni), plus a 7 keV thermal bremsstrahlung
component to account for the LMXBs (and the emission from the CXB that
is relatively insignificant in the inner region, and -- for the inner
region -- any activity associated with the dormant supermassive black
hole; see Loewenstein et al. 2001). The absorption is fixed at the
Galactic column density of $1.65~10^{20}$ cm$^{-2}$ \citep{dl90}. When
free to vary, the bremsstrahlung temperature and column density are
found to be consistent with these values; and, the {\bf vapec}
parameters essentially unchanged. This is also the case if a power-law
model is adopted to characterize the LMXBs. The inner region
bremsstrahlung flux is consistent with previous estimates of the LMXB
contribution \citep{ath07}, and the outer region bremsstrahlung flux
consistent with the expected sum of LMXB and CXB components. The Si
K-edge region (1.82-1.841 keV) was excluded from the spectra, although
this had negligible effect on the final fits aside from a small
reduction in $\chi^2$. A constant XIS1/XIS0+3 multiplicative factor is
included and generally ranges from 1.01--1.05. The \cite{gs98} solar
abundance standard is adopted. Since no elements with features of
significant strength are tied to solar ratios, the resulting
abundances may be re-scaled to other solar standards. Of course, we
use the same solar standard in the models that we compare with the
results of spectral fitting. C emission lines are outside of the {\it
 Suzaku} bandpass, so the C abundance is fixed at solar. The inner
and outer regions are separately fitted. The outer spectrum requires
an additional cooler component, presumably from GXB emission that we
model by a single temperature thermal plasma model with all abundances
except N and O fixed at solar (allowing other elements to vary, or
adding a second temperature component did not improve fits or affect
hot ISM parameters of interest). Because the GXB is more prominent
there, the outer spectra are used to determine the GXB parameters
(except for the normalization) that are then fixed in analyzing the
inner spectra. The relative fluxes of the GXB component in the inner
and outer regions are consistent with the ratio of their projected
areas, and the overall GXB surface brightness, temperature ($0.15\pm
0.04$ keV), and oxygen abundance (0.15, 0.11-0.29 at 90\% confidence)
are consistent with what was found in the source-free Ursa Minor Dwarf
Spheroidal Field \citep{lkb09} -- there is no evidence that any of the
``GXB'' constitutes an intrinsic NGC 4472 cool component. Fits to the
inner spectra require two ISM temperature components. The temperatures
we find in the inner region are $0.798\pm 0.004$ and $1.42\pm 0.05$
keV. The outer region may be fitted with a single-temperature (1.2
keV) model (reduced $\chi^2=1.05$), but is better fitted
($\Delta\chi^2=96$) with a two-temperature ($0.93\pm 0.08$ and
$1.55\pm 0.10$ keV) model. These are consistent with the temperature
structure we derive from spectra extracted over a finer series of
annuli (see next section). The best-fit N abundances are 0 with upper
limits of 0.75 (0.6) solar in spectral fits to the inner (outer)
spectra, and are fixed in the final fits. The addition of a third ISM
temperature component did not significantly affect the best-fit
abundances.

Spectra (with residuals to the best-fit model) and best-fit models,
are displayed in detail in Figures 1-4.  The best fit abundances and
90\% ($\Delta\chi^2=2.7$) confidence limits are shown in Table 1. The
best-fit abundance ratios, with respect to Fe, for the inner and outer
{\it Suzaku} spectra are compared in Figure 5. Note that O, Mg, Si, S,
Ar, Ca, and Al (K-) line features are well-resolved (though the latter
is weak), but that Ne (K-) and Ni (L-) line features are not, with
respect to the Fe L-line complex.

\begin{figure}
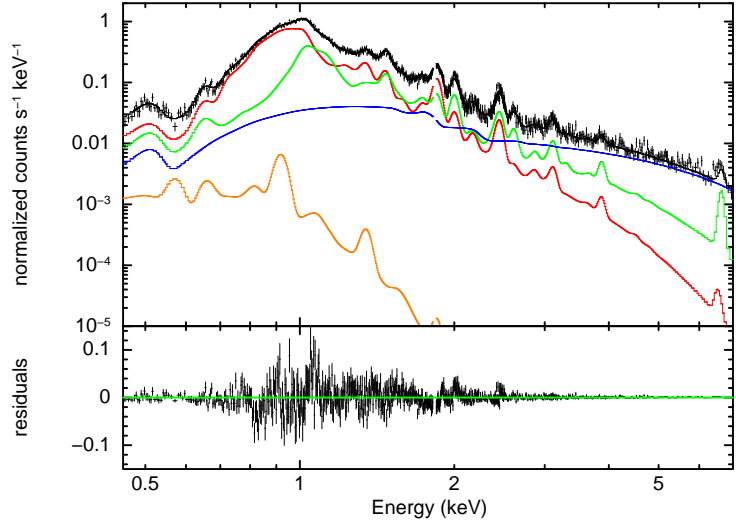

\centering \includegraphics[scale=0.4,angle=-90]{fig1a.eps}
\hfil
\caption{{\bf top (a)}: XIS0+3 inner region spectrum and best-fit
 model (black data points and histogram). Also separately shown are
 the GXB thermal plasma (orange), thermal bremsstrahlung LMXB (blue),
 and hotter (green) and cooler (red) hot ISM thermal plasma
 components. {\bf bottom (b)}: Same for XIS1.} \hfil \centering
\includegraphics[scale=0.4,angle=-90]{fig1b.eps}
\end{figure}

\begin{figure}
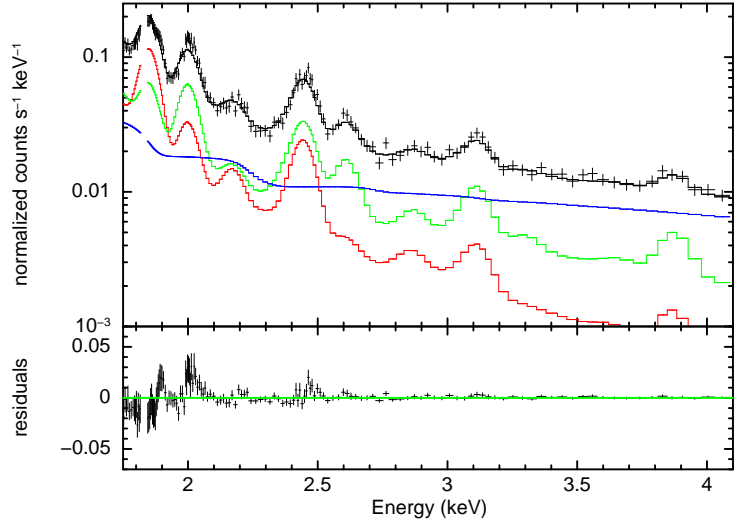

\centering
\includegraphics[scale=0.4,angle=-90]{fig2a.eps}
\hfil
\caption{Same as Figure 1, zoomed to the Si/S/Ar/Ca spectral
region. Prominent features from left to right predominantly originate
in the following ions: Si XIII ($K\alpha$), Si XIV ($K\alpha$), Si XII
($K\beta$), S XV ($K\alpha$), S XVI ($K\alpha$), S XV ($K\beta$), Ar XVII
($K\alpha$), Ca XIX ($K\alpha$).} \hfil \centering
\includegraphics[scale=0.4,angle=-90]{fig2b.eps}
\end{figure}

\begin{figure}
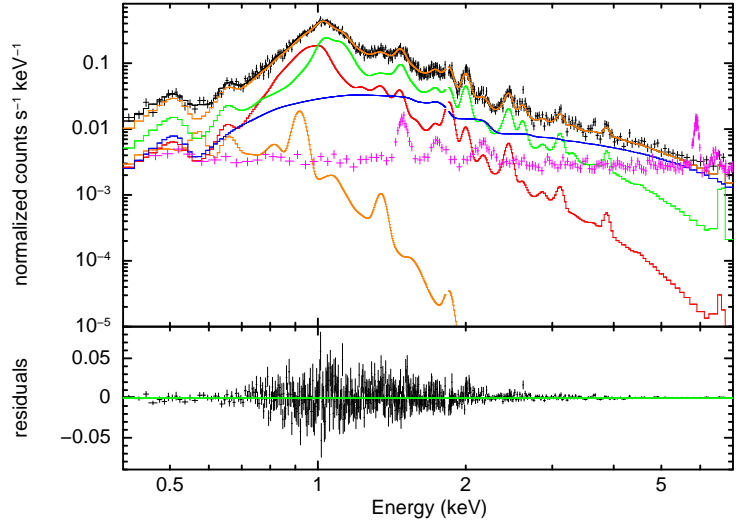

\centering
\includegraphics[scale=0.4,angle=-90]{fig3a.eps}
\hfil
\caption{Same as Figure 1 for outer region spectra, with NXB (purple)
 contribution and sum of non-background components (orange) now
 included.} \hfil \centering
\includegraphics[scale=0.4,angle=-90]{fig3b.eps}
\end{figure}

\begin{figure}
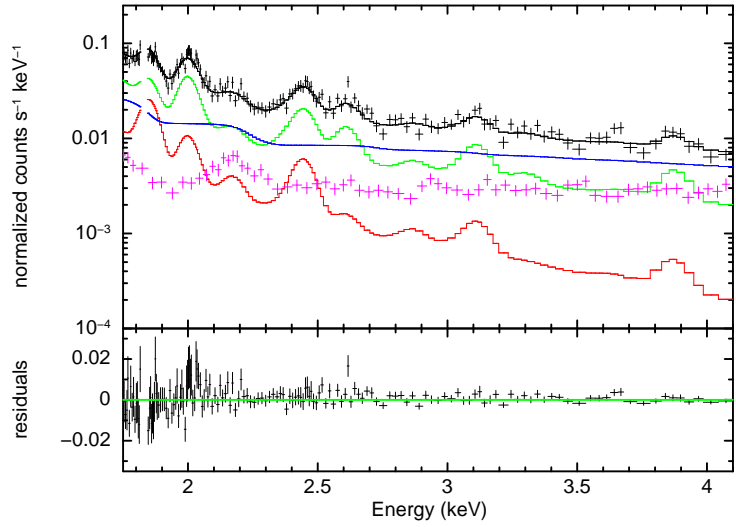

\centering
\includegraphics[scale=0.4,angle=-90]{fig4a.eps}
\hfil
\caption{Si/S/Ar/Ca spectral region as in Figure 2 for outer spatial
 region, including NXB (purple) contribution.} \hfil \centering
\includegraphics[scale=0.4,angle=-90]{fig4b.eps}
\end{figure}

\begin{figure}
\plotone{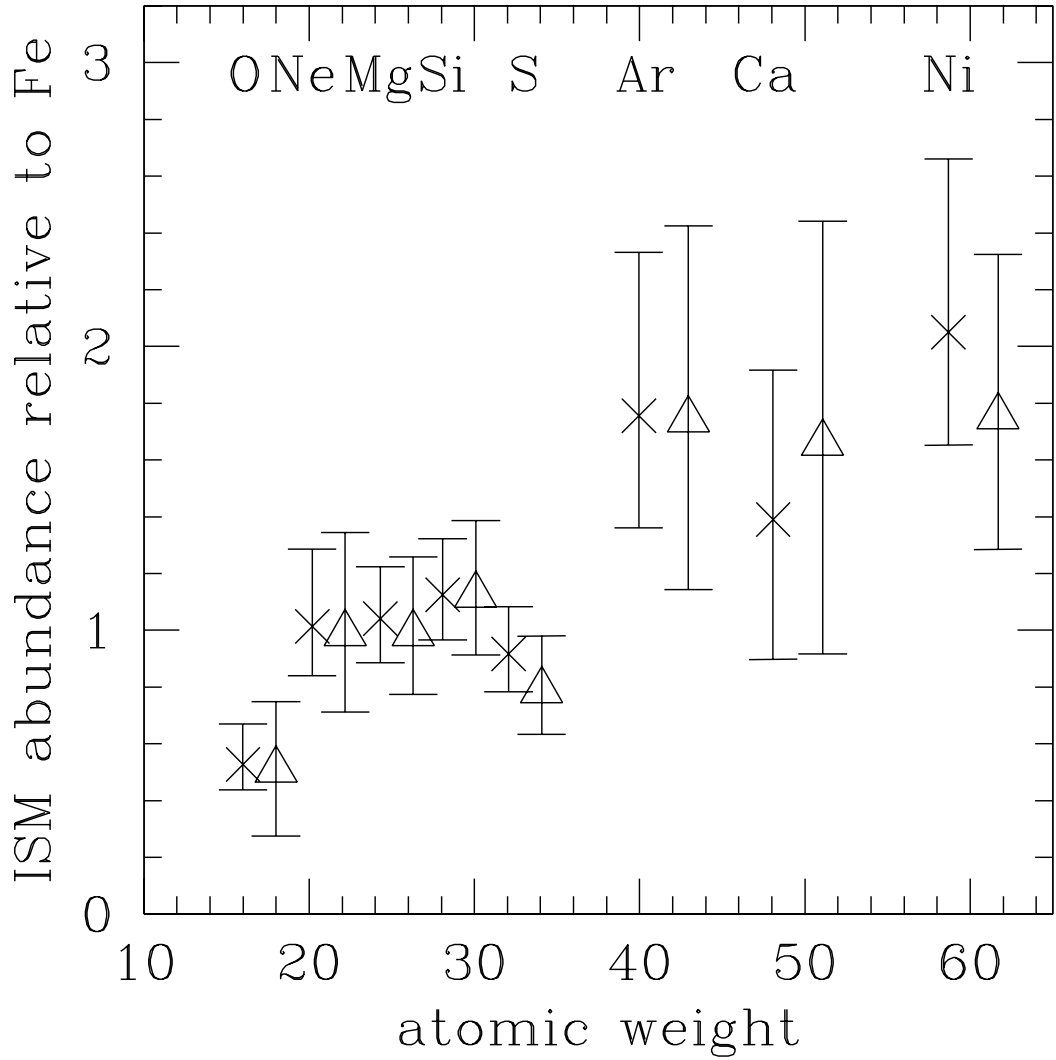}
\caption{Inner (crosses), and outer (open triangles, offset to right)
{\it Suzaku} abundance patterns. Ca is shifted to the right.}
\end{figure}

\subsection{\it XMM-Newton Spectral Analysis and Comparison to with
{\it Suzaku}}

{\it XMM-Newton} spectra are analyzed using the same models that we
applied to the {\it Suzaku} analysis. Fits to the {\it XMM-Newton}
EPIC MOS1+2 and pn data in the same extraction regions yielded best
fits that are generally consistent with the above, though with some
discrepancies (Table 1, Figure 6). Although blank sky background is
subtracted from these spectra, an additional small GXB component is
included. The best-fit GXB normalization is $3-10\times$ lower than in
the (non-GXB-subtracted) {\it Suzaku} spectra. The effect of this
addition is to improve the spectral fits ($\Delta\chi^2=53$ for the
MOS, 16 for the pn) and increase the parameter uncertainties. As with
the {\it Suzaku} spectra, a two-temperature plasma is required for the
hot ISM.

There are substantial differences in the derived best-fit abundances
of Ar, Ca, and Ni derived from EPIC and {\it Suzaku} spectra (Figure
6a); but, the errors are large (particularly for the former) and the
90\% confidence ranges overlap. The most serious discrepancies are the
low outer-region Mg abundance and inner-region O abundance derived
from EPIC spectra.

The {\it Suzaku} abundance pattern we derive in the inner region is in
very good accord with that found by Ji et al. (2009) from analysis of
EPIC and RGS data, and from our own RGS analysis (Figure 6a, Table 1).
Although the latter samples a smaller region, the best-fit
temperatures and abundances are consistent with our analysis of the
inner region {\it Suzaku} data except for N (best RGS fit: $3.7\pm
1.7$ at 90\% confidence) that is sensitive to the treatment of the
background, and (marginally) Mg.

The {\it XMM-Newton} EPIC angular resolution allows us to determine
the temperature and abundances in 9 annular rings out to
300$\arcsec$. In our analysis we find that a two-temperature {\bf
vapec} plasma model is needed for most of the radial bins. This
requirement is driven by residuals to one-temperature fits in the CCD
spectra near the energy of the Fe XVII line. This is in contrast to
the analysis of \cite{nm09}, who adopt a one-temperature model in
deriving a {\it deprojected} temperature profile. In our analysis we
have used fairly wide annuli (30$\arcsec$ minimum) and adjusted their
widths to include at least 20000 counts in order to allow us to
accurately determine the abundances in each ring. This requirement
results in extraction regions that span a larger fraction of the
temperature gradient so that two-temperature models are needed to fit
the data. Based on a comparison with one temperature models in the
innermost annulus, the two temperature model is required at the
99.99\% level. All other annuli require a two temperature fit with at
least a 90\% confidence, except for the outermost ring where the two
temperature fit is only significant at the 80\% level. We adopt the
two-temperature model for this annulus as well to facilitate
comparison with the other annuli. Our temperature profile is shown in
Figure 7, and shows a positive gradient. The cooler component
temperature is about 60\% that of the hotter at all radii. The ratio
of hot to cool components increases with radius, so that in the outer
annuli the hotter component dominates the emission.

Utilizing these same annular extraction regions, we search for
abundance gradients on a finer scale than that of the {\it Suzaku}
data for the best-determined abundances: O, Ne, Mg, Si, S, and Fe. We
determine the abundance gradient for each element with respect to Fe;
and, in general the profile is consistent with a constant. The only
ratio that shows a gradient is Mg-to-Fe, which is flat in the inner
150 - 200$\arcsec$ and declines to about half the central value at
$\sim$ 300$\arcsec$.

\begin{deluxetable}{ccccccccccccc}
\tabletypesize{\scriptsize}
\tablewidth{0pt}
\tablecaption{Best Fit ISM Abundances}
\tablehead{\colhead{} & \colhead{$\chi^2_{\nu}$} &
\colhead{0} & \colhead{Ne} & \colhead{Mg} & \colhead{Al} &
\colhead{Si} & \colhead{S} & \colhead{Ar} & \colhead{Ca} &
\colhead{Fe} & \colhead{Ni}}
\startdata 

{\bf S-in} & 1.13 & $1.08_{0.15}^{0.24}$ & $2.08_{0.30}^{0.46}$ &
$2.14_{0.25}^{0.17}$ & $3.98_{0.97}^{1.46}$ & $2.31_{0.25}^{0.18}$ &
$1.88_{0.21}^{0.18}$ & $3.61_{0.74}^{1.04}$ & $2.86_{0.98}^{0.99}$ &
$2.05_{0.19}^{0.32}$ & $4.21_{0.71}^{1.07}$\\

{\bf MOS-in} & 1.18 & $0.77_{0.42}^{0.54}$ & $1.64_{0.32}^{0.43}$ &
$2.08_{0.29}^{0.27}$ & $7.35_{1.50}^{2.10}$ & $2.21_{0.17}^{0.40}$ &
$1.85_{0.20}^{0.27}$ & $2.85_{1.04}^{1.05}$ & $2.38_{1.53}^{1.54}$ &
$2.13_{0.19}^{0.31}$ & $4.31_{0.83}^{1.13}$\\

{\bf pn-in} & 1.07 & $0.31_{0.29}^{0.32}$ & $1.13_{0.33}^{0.38}$ &
$1.59_{0.28}^{0.17}$ & $4.15_{1.30}^{1.63}$ & $2.00_{0.16}^{0.32}$ &
$1.53_{0.18}^{0.20}$ & $1.73_{1.05}^{1.16}$ & $0.40_{0.40}^{1.84}$ &
$1.75_{0.13}^{0.21}$ & $6.22_{1.21}^{1.57}$\\

{\bf EPIC-in} & 1.18 & $0_{0}^{0.27}$ & $1.50_{0.22}^{0.24}$ &
$1.98_{0.16}^{0.17}$ & $6.48_{1.29}^{0.83}$ & $2.26_{0.15}^{0.18}$ &
$1.80_{0.18}^{0.13}$ & $2.42_{0.80}^{0.80}$ & $0.92_{0.92}^{1.10}$ &
$2.07_{0.18}^{0.16}$ & $5.66_{0.86}^{0.64}$\\

{\bf RGS} & 1.16 & $0.81_{0.15}^{0.16}$ & $2.05_{1.01}^{0.52}$ &
$0.97_{0.76}^{0.76}$ & \nodata & \nodata & \nodata & \nodata &
\nodata & $2.42_{0.19}^{0.27}$& \nodata\\

{\bf Suz-out} & 0.99 & $0.75_{0.33}^{0.33}$ & $1.45_{0.36}^{0.47}$ &
$1.45_{0.26}^{0.32}$ & $2.41_{1.30}^{1.52}$ & $1.65_{0.23}^{0.29}$ &
$1.16_{0.17}^{0.21}$ & $2.56_{0.82}^{0.92}$ & $2.44_{1.05}^{1.08}$ &
$1.47_{0.19}^{0.23}$ & $2.58_{0.60}^{0.73}$\\

{\bf MOS-out} & 1.07 & $1.32_{0.59}^{0.87}$ & $1.12_{1.07}^{1.24}$ &
$0.59_{0.59}^{0.76}$ & \nodata & $1.79_{0.44}^{0.67}$ &
$1.49_{0.47}^{0.64}$ & $4.42_{3.38}^{3.90}$ & $6.22_{5.08}^{5.64}$ &
$1.55_{0.38}^{0.56}$ & $0.97_{0.97}^{1.64}$\\

{\bf pn-out} & 1.04 & $0.30_{0.30}^{0.56}$ & $0_{0}^{1.35}$ &
$0_{0}^{0.33}$ & \nodata & $1.07_{0.32}^{0.51}$ & $1.01_{0.34}^{0.44}$
& $1.89_{1.89}^{3.11}$ & $2.23_{2.23}^{4.47}$ & $1.22_{0.37}^{0.63}$ &
$1.67_{0.68}^{1.71}$\\

{\bf EPIC-out} & 1.06 & $0.58_{0.30}^{0.45}$ & $0.58_{0.58}^{0.82}$ &
$0.11_{0.11}^{0.36}$ & \nodata & $1.28_{0.23}^{0.38}$ &
$1.13_{0.28}^{0.30}$ & $3.04_{2.05}^{2.10}$ & $3.57_{3.04}^{3.08}$ &
$1.27_{0.23}^{0.40}$ & $1.16_{0.86}^{1.01}$\\

\enddata 

\tablecomments{Reduced $\chi^2$ and best fit abundances relative to
the Grevesse \& Sauval (1998) solar standard, as well as 90\%
confidence uncertainties for the inner region {\it Suzaku} (XIS0+3 and
XIS1) spectra ({\bf Suz-in}), outer region {\it Suzaku} spectra ({\bf
Suz-out}), the inner and outer region MOS1+2 and pn spectra ({\bf
MOS-in}, {\bf MOS-out}, {\bf pn-in}, {\bf pn-out}), the RGS-1\&2
spectra ({\bf RGS}), and the joint fits to EPIC spectra ({\bf
EPIC-in}, {\bf EPIC-out}).}
\end{deluxetable}

\begin{figure}
\plottwo{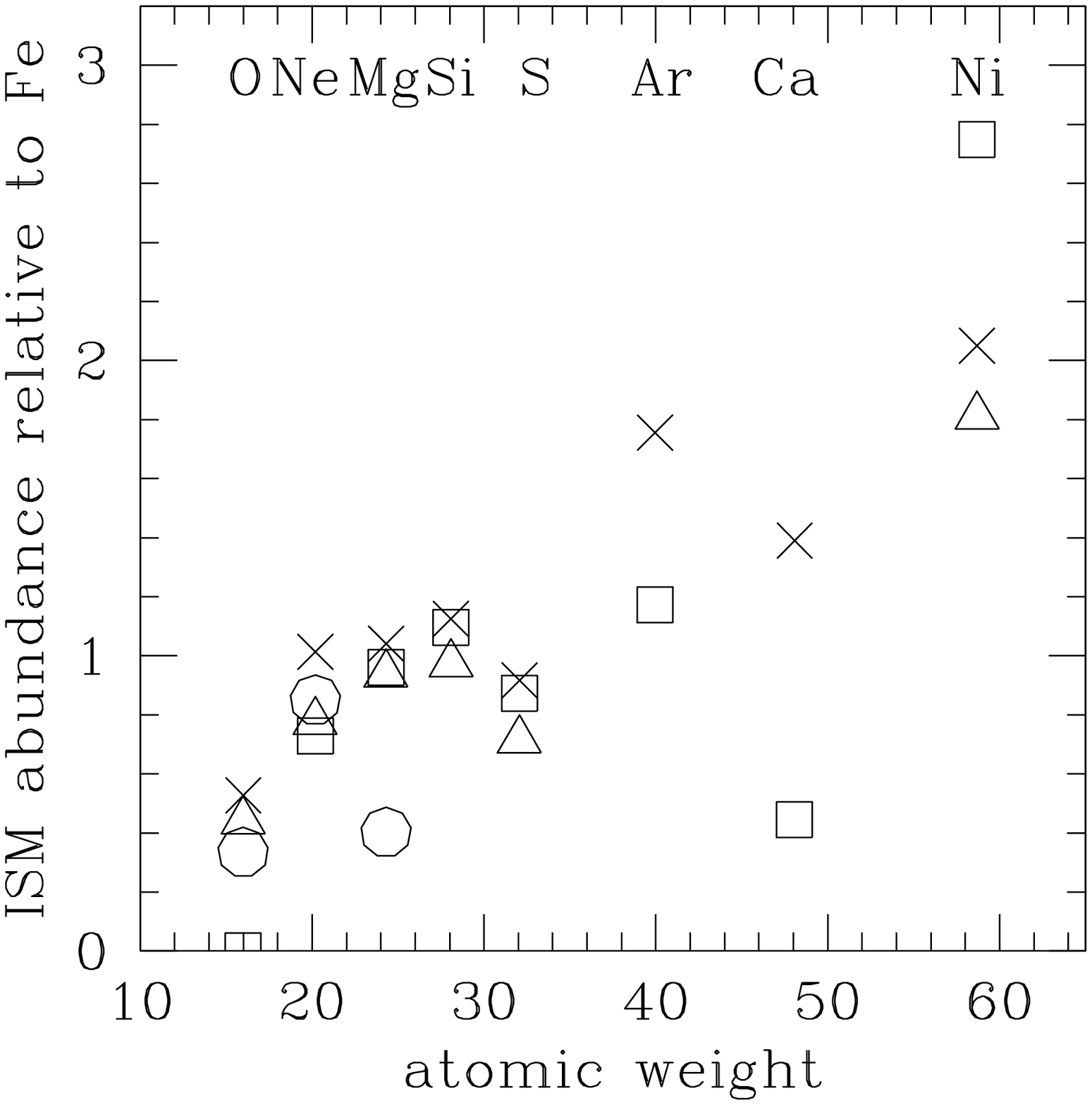}{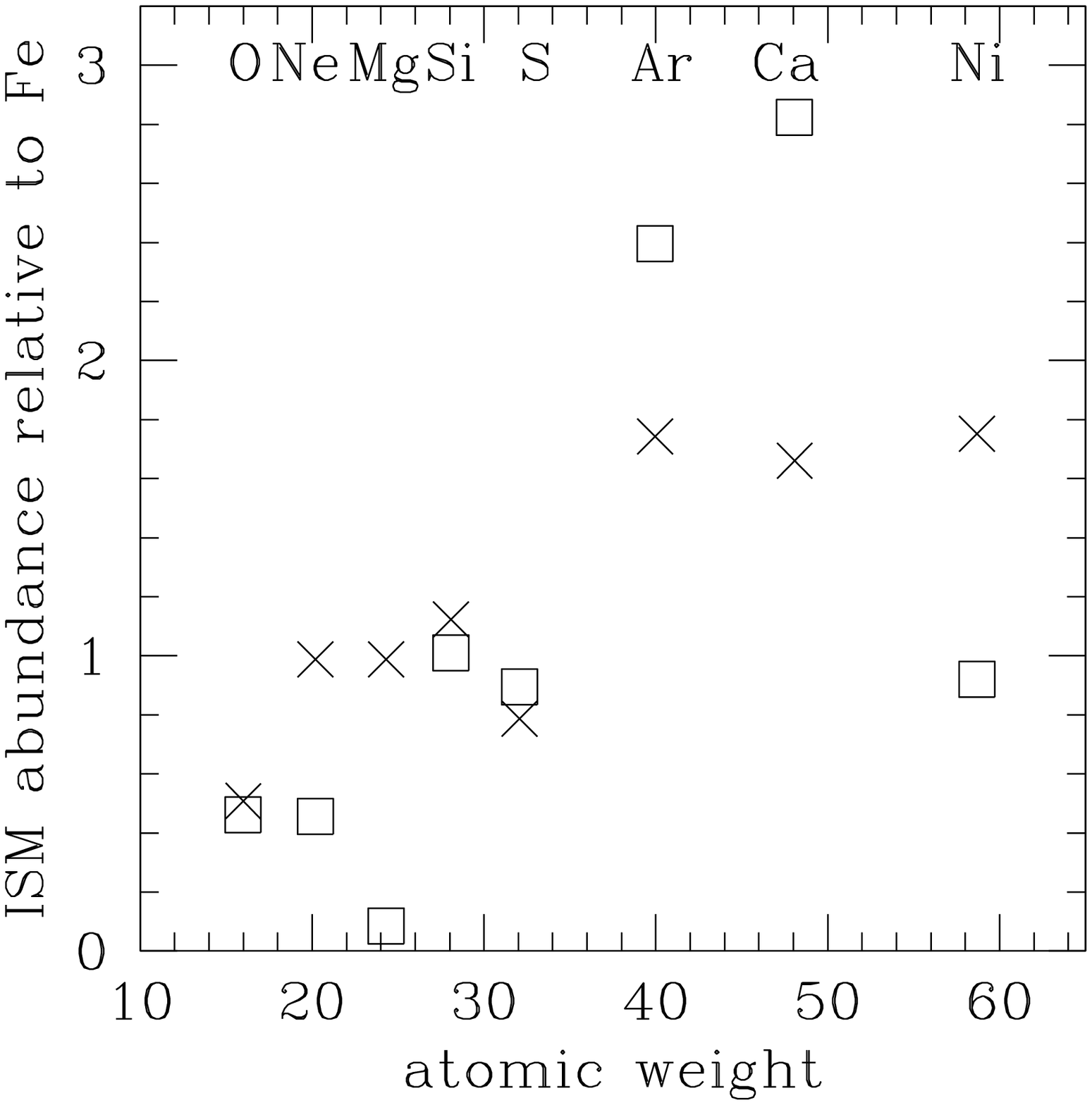}
\caption{{\bf left (a):} Inner region abundance pattern (normalized to
Fe) derived from {\it Suzaku} XIS0+3 and XIS1 (crosses), {\it
XMM-Newton} EPIC from our analysis (open squares) and from
\citet{ji09} (open triangles), and {\it XMM-Newton} RGS (open
polygons). {\bf right (b):} Outer region abundance pattern (normalized
to Fe) derived from {\it Suzaku} XIS0+3 and XIS1 (crosses), and {\it
XMM-Newton} EPIC (open squares).}.
\end{figure}

\begin{figure}
\centering
\includegraphics[scale=0.4,angle=0]{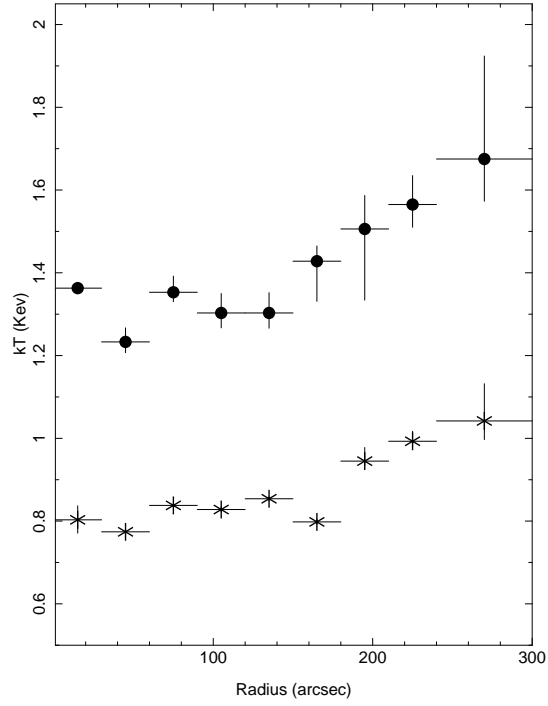}
\hfil
\caption{The projected temperature profile for N4472 from the center
out to 300$\arcsec$. The crosses denote the low temperature component,
and the filled circles the hotter component. The error bars show the
90\% confidence range. The horizontal bars denotes the radial range
over which the ISM spectral parameters were derived.} \hfil
\centering
\end{figure}

In the following sections, we adopt the abundances that we infer from
fitting the {\it Suzaku} spectra for comparison with simple chemical
evolution models. We consider these more robust, for the crude spatial
binning that we employ, because of the lower background and sharper
energy resolution of {\it Suzaku}. This is illustrated in Figure 8
that compares MOS1, pn, and XIS0 inner region spectra, and Figure 9
that compares pn and XIS0 outer region spectra. The {\it Suzaku}
spectral features, including the 1.0 keV Ne X bump embedded in the
Fe-L line complex (note, also, the excellent agreement of the {\it
 Suzaku} inner spectrum Ne abundance with that derived from the RGS
spectrum where this feature is resolved), the OVII line at 0.65 keV,
and the Mg XI and Mg XII lines at 1.35 and 1.47 keV are significantly
sharper than their EPIC counterparts (and in some instances sharper
than the resolution in the latest {\it Suzaku} calibration). The
reduced high energy background is particularly significant in the
Si/S/Ar/Ca spectral regime for the outer region (Figure 8).

We note the formal measurement of a large Al abundance in the inner
region (with {\it XMM-Newton} and {\it Suzaku}) and the outer regions
({\it Suzaku} only -- there was residual non-X-ray Al fluorescence
emission in the outer-region MOS spectra). This would be the first
detection of an odd-atomic-number element in the hot ISM of an
elliptical galaxy. An Al abundance tied at solar results in a poorer
fit over the 1.5-1.8 keV region ($\Delta\chi^2=29$ for the inner {\it
Suzaku} spectrum). However, the inferred 1.6 keV Al XII and 1.7 keV Al
XIII features are weak, and possibly sensitive to uncertainties in the
characterization of the continuum in their vicinity. We do not include
Al abundances in the model constraints discussed in the subsequent
sections, but do sometimes display them in comparisons of models with
data.

\begin{figure}
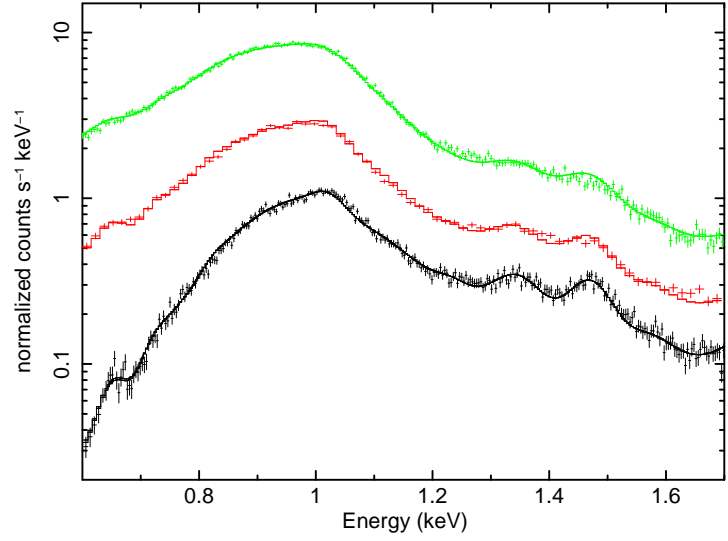

\centering
\includegraphics[scale=0.4,angle=-90]{fig8a.eps}
\hfil
\caption{{\bf top (a)}: MOS1 (red), pn (green), and XIS0 (black) inner
region spectra in the 0.6--1.7 keV bandpass. {\bf bottom (b)}: Same for
1.75-4 keV bandpass.}
\hfil 
\centering
\includegraphics[scale=0.4,angle=-90]{fig8b.eps}
\end{figure}

\begin{figure}
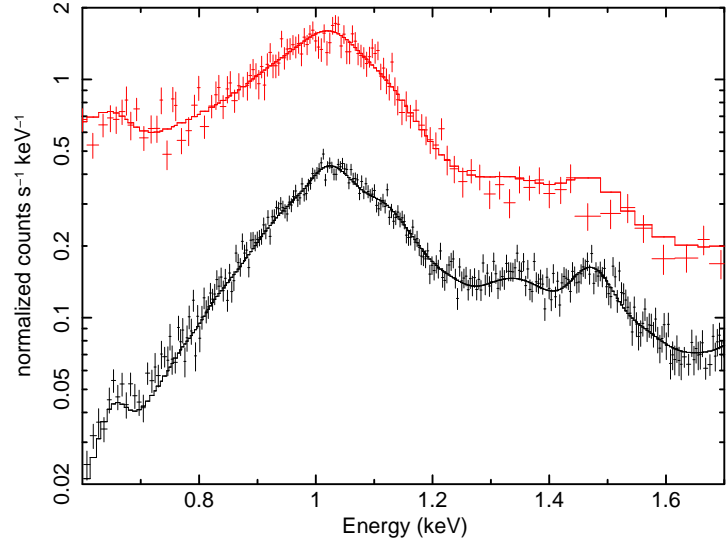

\centering
\includegraphics[scale=0.4,angle=-90]{fig9a.eps}
\hfil
\caption{{\bf top (a)}: pn (red), and XIS0 (black) outer region
spectra in the 0.6--1.7 keV bandpass. {\bf bottom (b)}: Same for
1.75-4 keV bandpass.}
\hfil 
\centering
\includegraphics[scale=0.4,angle=-90]{fig9b.eps}
\end{figure}

\section{Interpreting the NGC 4472 Abundance Pattern}

Our method of interpretation, described in more technical detail in
Appendix A, may be summarized as follows. We assume that the ISM
abundances are determined following the cessation of the main era of
star formation $<9$ Gyr ago for NGC 4472 \citep{hb06,gal08}, at which
time the stellar abundances are established. The internal sources of
mass and metals are stellar mass loss -- calculated using remnant
masses and main sequence turnoff times from the literature -- and SNIa
explosions. Effects of outflow (including the possibility of a
mismatch between the metallicity of the gas in the galaxy and that
flowing out of the galaxy), initial conditions, and different time
dependencies for the mass return and SNIa rates are considered. For a
set of simple one-zone models (``closed-box'', and ``tracking''
solutions where mass return and SNIa have the same time-dependence --
steady state solutions are a subset of the latter), we solve the
equations expressing the conservation of overall ISM mass and mass of
the $ith$ element for the present-day ISM abundances. For these simple
models, solutions generally are of the form
\begin{equation}
f^i_{\rm ISM}=rf^i_{\rm stars}+sqy_{\rm SNIa}^i,
\end{equation}
where $f^i_{\rm ISM}$ and $f^i_{\rm stars}$ are the $ith$ element ISM
and stellar mass fractions, respectively; $r$ is a stellar mass loss
dilution ($r<1$) or enhancement ($r>1$) factor, and $sq$ is a term
that determines the importance of SNIa enrichment relative to that
from stellar mass loss. As explained in Sections A1-3, $q$ is the
ratio of present-day SNIa to mass return rate, $q=0.00667\theta_{\rm
SNIa}/\theta_{\rm MR}~M_{\odot}^{-1}$, where $\theta_{\rm MR}$ is the
specific mass loss rate in units of $2.4~10^{-11}~M_{\odot} {\rm
yr}^{-1}~L_{B\odot}^{-1}$ (we henceforth assume $\theta_{\rm
MR}=1$), and $\theta_{\rm SNIa}$ is the specific SNIa rate in units of
0.16 SNU \citep{cap97}. The dimensionless $r$ and $s$ parameters may
be element-dependent, but we do not consider that generalization here.
The input parameters are (1) the stellar $[\alpha/Fe]$ ratio that we
use to determine the stellar abundance pattern using (IMF-weighted,
solar metallicity progenitor) SNII+HN (hypernova) and SNIa yields from
\cite{nom97,kob06}, (2) the ratios of the SNIa yields for each element
relative to that of Fe, (3) the ratio of ISM Fe abundance to stellar
Fe abundance, and (4) the dilution parameter $r$. The output is the
ISM abundance pattern relative to Fe. Specifying the ISM and stellar
Fe abundances and SNIa yields (not just the ratios), determines $sq$
and the predicted values of the ISM abundances. The $r$ and $sq$
parameters can then be interpreted in terms of relative SNIa
frequency, outflow parameters, and initial condition in the context of
different models (Section A3). We generally consider $[\alpha/Fe]_{\rm
stars}=0.25, 0.15, 0, -0.2$ (where the sum used to compute $\alpha$
includes only those elements measured in the ISM), an ISM/stellar Fe
abundance ratio 1.65 based on a gas-mass weighted average of our inner
and outer region {\it Suzaku} abundances and the stellar Fe abundance
from \cite{hb06}, and $r=1, 2/3, 1/3$.

A value of $r<1$ may be explained in several ways (Sections A.1-A.3).
In the steady-state models we consider, $r=1/3$ (2/3) corresponds to
an outflow bias of 2.0 (0.5) -- i.e, that gas is outflowing with
$1.5\times$ ($3.0\times$) the mean metallicity inside the galaxy;
$r=1$ corresponds to no such bias. For the closed box models with the
extreme initial condition of 0 initial ISM metallicity, $r=1/3$ (2/3)
corresponds to having two-thirds (one-third) of the present-day ISM
mass at 84\% (91\%) of the Hubble time. The dilution is attributed to
this preexisting low-metallicity gas; $r=1$ corresponds to (the
reasonable) initial conditions where the ISM and stellar abundances
are equal. For tracking solutions, where winds are driven by SNIa and
the mass return and SNIa rates have identical time-dependencies, we can
also consider the extreme case where the ISM metallicity is 0 at the
initial time of $0.7$ of the Hubble time. Here, $r=1/3$ (2/3)
corresponds to having $1.8\times$ ($35\times$) the current ISM
mass. The dilution is due to the large initial reservoir of
low-metallicity gas that is driven out of the galaxy while being
replenished by SNIa-enriched stellar mass return; $r=1$ again
corresponds to (the reasonable) initial conditions where the ISM and
stellar abundances are equal.

We consider two sets of SNIa yields from \cite{nom97} -- the standard
W7 deflagration model, and the WDD1 delayed detonation model that
represents the most radical departure from W7 in terms of
nucleosynthetic yields. A comparison of the {\it Suzaku}-derived ISM
abundance pattern inside the optical galaxy (the gas-mass weighted sum
of our inner and outer abundances) and the implied stellar abundance
pattern is shown in Figure 10. For stellar abundances derived using the
W7 yields, the ISM abundances are similar to those in the stars
assuming a solar or slightly supersolar $[\alpha/Fe]_{\rm stars}$ --
with the notable exceptions of ISM overabundances of Ar and Ca and an
underabundance of O (all relative to Fe). For the WDD1 yields, the ISM
again is similar to the $[\alpha/Fe]_{\rm stars}\sim 0$ pattern, but
now Ni is overabundant while O and S are underabundant in the
ISM. Since direct SNIa enrichment is expected to distort the ISM
pattern relative to that in the stars, these comparisons should be
considered referential; however, some of these anomalies persist.

\begin{figure}
\plottwo{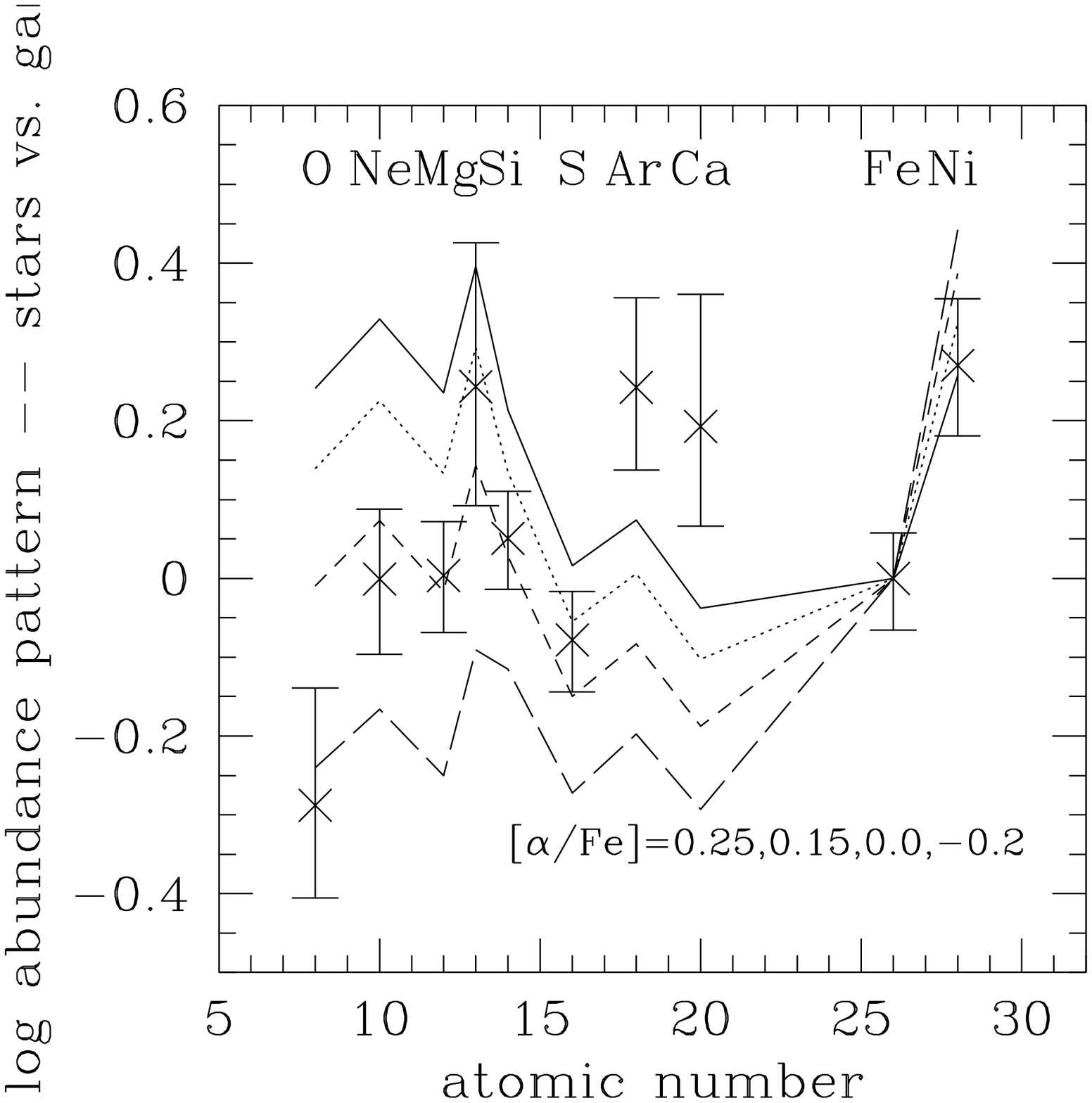}{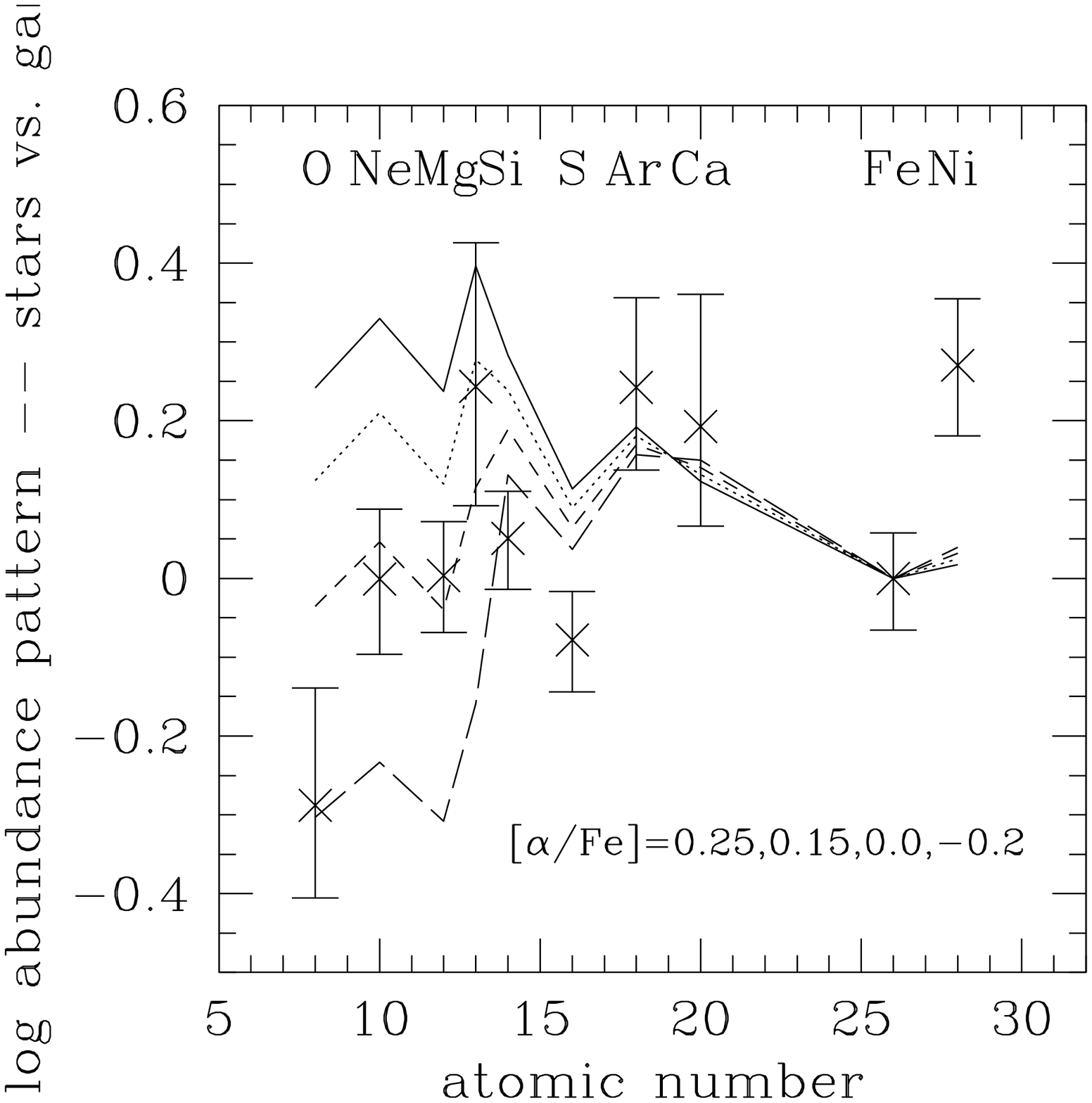}
\caption{Comparison of abundances (including Al -- located between Mg
and S), relative to Fe, in the ISM (crosses with errorbars) and that
expected in the stars for the cases where W7 SNIa {(\bf a, left)} and
WDD1 {(\bf b, right)} SNIa yields \citep{nom97} are used in
conjunction with SNII+HN yields from \cite{kob06} to set the stellar
abundance pattern for a given $[\alpha/Fe]_{\rm stars}$. Shown are
$[\alpha/Fe]_{\rm stars}$=0.25 (solid line), 0.15 (dotted), 0.0
(short-dashed) and -0.2 (long-dashed).}.
\end{figure}

To review how we compare these models with the observed abundances,
sufficient SNIa enrichment (value of $sq$) is added to the diluted
(since we consider $r\le 1$) stellar Fe abundance to match the
observed ISM Fe abundance. The SNIa rate corresponding to ``sufficient
enrichment'' is model-dependent -- but only weakly so for quiescent
systems, where $s\approx r$, i.e. $q\approx (sq)/r$ or $\theta_{\rm
SNIa}\approx 150 (sq)/r$. The magnitudes of the stellar mass loss and
SNIa terms are now set and may be used, in conjunction with the
stellar abundance pattern and SNIa yields, to determine the entire ISM
pattern. Figure 11 compares model predictions of the ISM abundances
with the observations for the case where W7 yields are adopted;
$\theta_{\rm SNIa}$ corresponding to $s=r$ is indicated.\footnote{We
note here that our results are not sensitive to our adoption of the
\cite{gs98} solar abundance standard, since all quantities are scaled
on this basis. Adoption of a different standard would shift observed
and model abundances in lockstep, although the value of
$[\alpha/Fe]_{\rm stars}$ corresponding to a given model would shift
somewhat.} The abundances of Ne, Mg, Si, S and Ni (as well as Al) are
well-explained as a combination of undiluted ($r=1$) stellar mass loss
from stars with $[\alpha/Fe]_{\rm stars}$=0.25 and SNIa enrichment at
a rate approximately one-sixth the standard early-type galaxy value,
$\theta_{\rm SNIa}=0.17$ (Figure 11a). However, Ar and Ca are
underpredicted by factors of order 2 and O overpredicted by a similar
factor. On the other hand the model with sufficient dilution to imply
$\theta_{\rm SNIa}=1$ for a steady-state model ($r=1/3$), the expected
low ratios with respect to Fe of Ne, Mg, Si, S, Ar, Ca are greatly at
odds with the observations; and, the Ni/Fe ratio is overpredicted
(Figure 11b). In the limit where the stellar abundances are purely
determined by SNII ($[\alpha/Fe]_{\rm stars}\sim 0.43$), the ISM
O-to-Fe ratio is reproduced but all other ratios with respect to Fe
are not (Figure 11b).

\begin{figure}
\plottwo{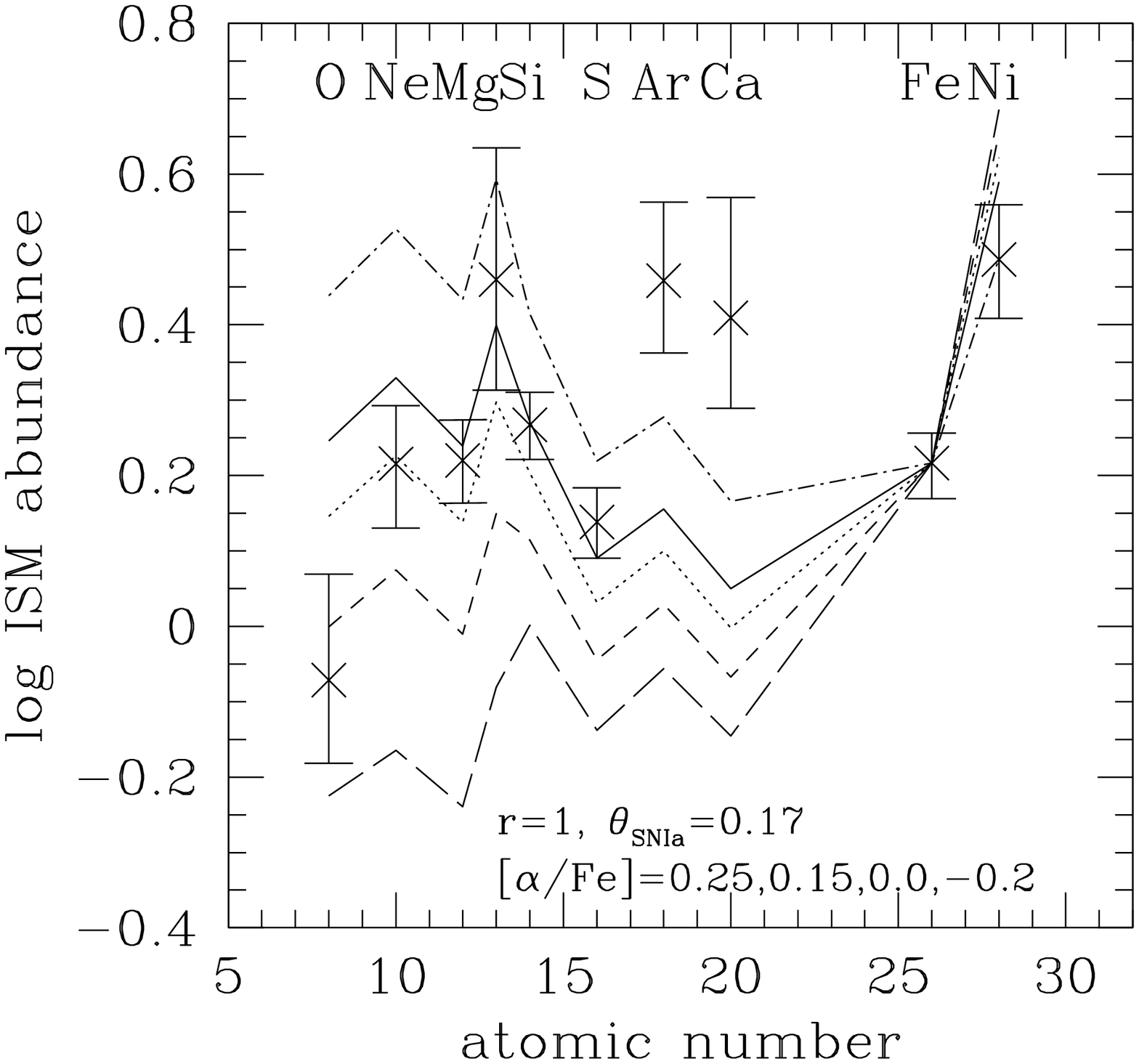}{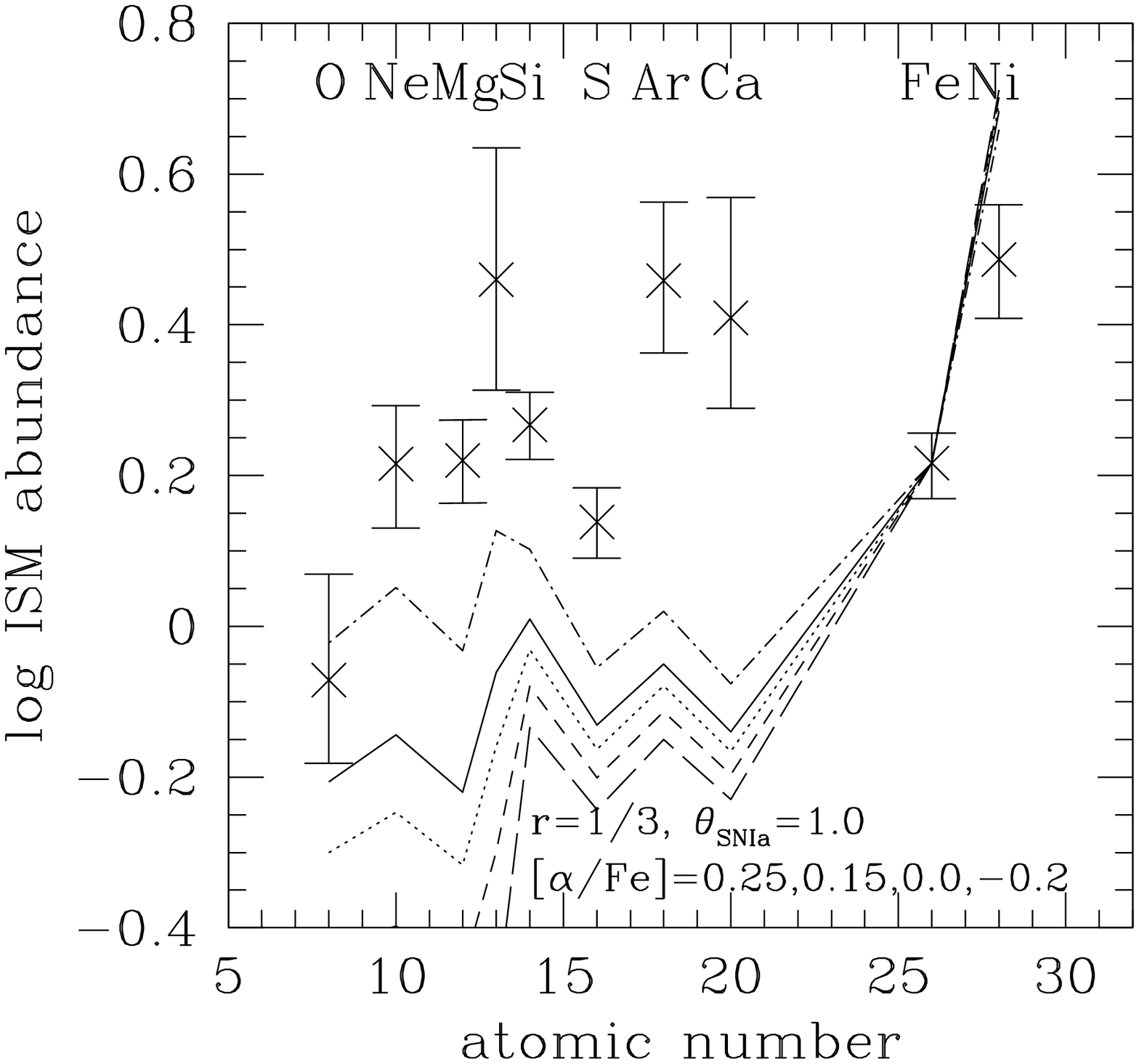}
\caption{Comparison of abundances (including Al -- located between Mg
and S), relative to Fe, in the ISM (crosses with errorbars) and those
predicted in models where W7 SNIa yields are adopted for $r=1$ {(\bf
left a:}) and $r=1/3$ {(\bf right b:}). Shown are $[\alpha/Fe]_{\rm
stars}$=0.25 (solid line), 0.15 (dotted), 0.0 (short-dashed) and -0.2
(long-dashed). Additionally, the $[\alpha/Fe]_{\rm stars}=0.45$ (pure
SNII stellar abundances) model is shown (dot-dashed line).}
\end{figure}

The WDD1 models, where SNIa produce more Si, S, Ar, and Ca and less Fe
and Ni, $r=1$ models with supersolar $[\alpha/Fe]_{\rm stars}$ now
adequately explain the ratios, with respect to Fe, of Ne, Mg, Ar, and
Ca (as well as Al) but overpredict Si (slightly) and S as well as 0
(Figure 12a). Ni is underpredicted by $\sim 2$, as it is for the
$r=1/3$ model. In the latter (Figure 12b), Si and O are matched by
models with $[\alpha/Fe]_{\rm stars}$ corresponding to a pure SNII
origin, but Ne and Mg are now underpredicted.

\begin{figure}
\plottwo{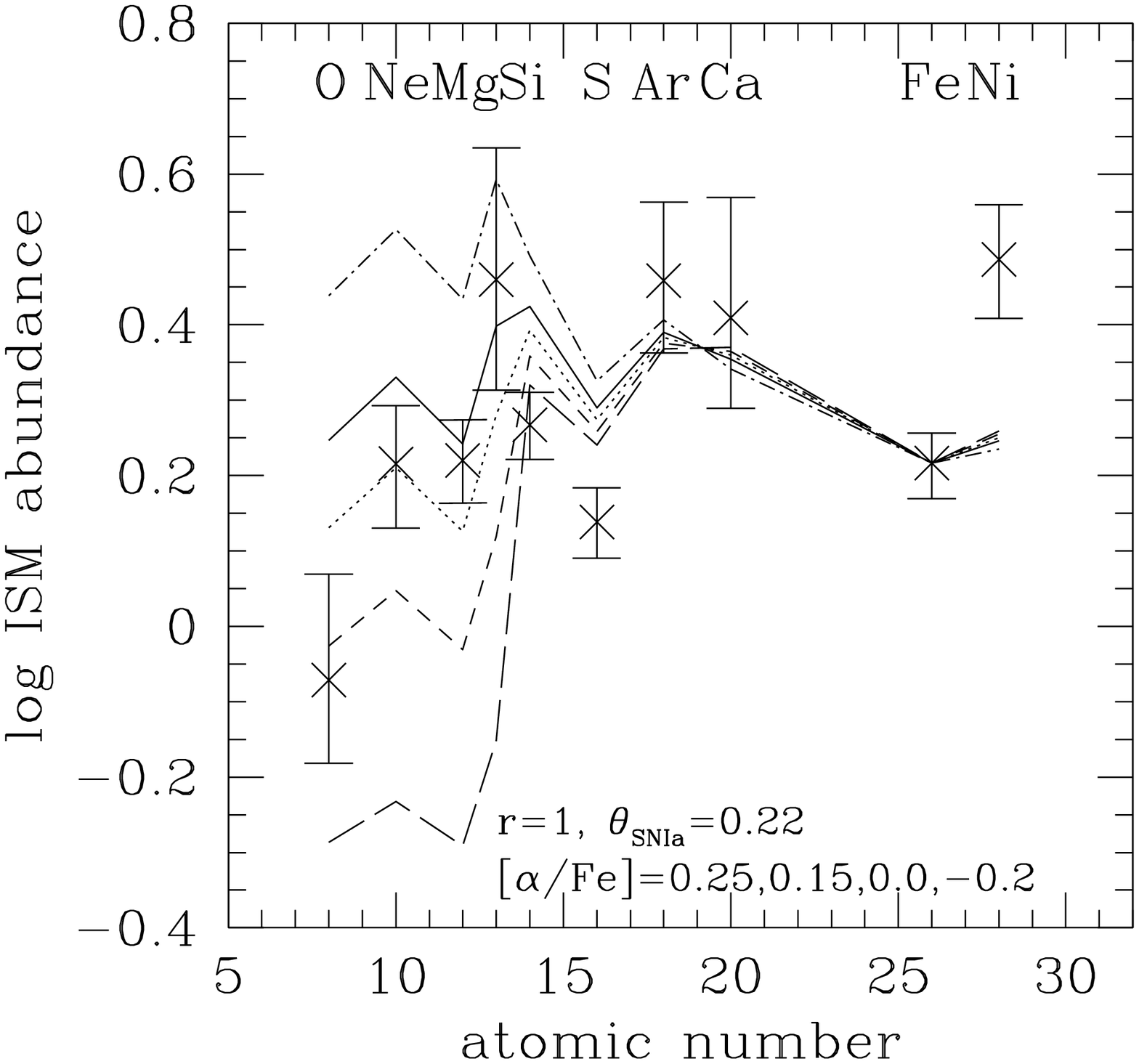}{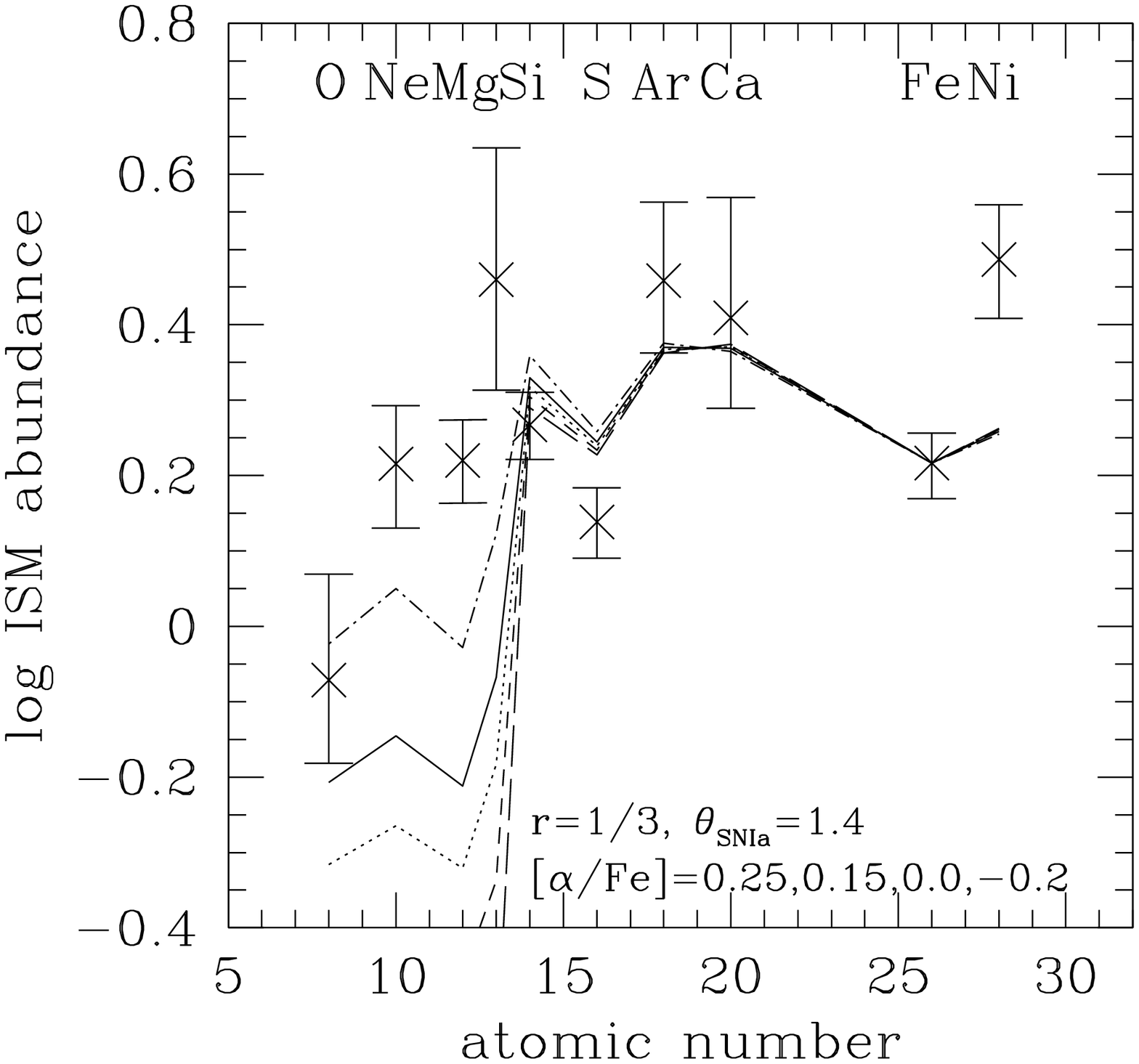}
\caption{Same as Figure 11 for WDD1 SNIa yields.}
\end{figure}

\section{Discussion}

In the most natural models -- whether they resemble the closed box,
steady state, or slightly more complex outflow models considered here
-- abundances (expressed as mass fractions) in the hot ISM of
elliptical galaxies should be well-approximated by
\begin{equation}
f^i_{\rm ISM}(1)=f^i_{\rm stars}+0.00667\theta_{\rm SNIa}y_{\rm SNIa}^i.
\end{equation}
Figures 11 and 12 clearly show that the modest overabundance of
interstellar Fe with respect to that in the stars, and the nearly
solar abundance ratios of Ne, Mg, Si, and S with respect to Fe in NGC
4472, are consistent with this expectation provided that $\theta_{\rm
SNIa}\sim 0.2$ and $[\alpha/Fe]_{\rm stars}\sim 0.2$. We thus confirm
the inference, based on optical data, that the stars in elliptical
galaxies have supersolar $[\alpha/Fe]$ ratios -- and infer that this
extends globally. The relative Fe abundances may be accommodated in
models with $\theta_{\rm SNIa}\sim 1$ only by introducing mechanisms
such as preferential outflow of metal rich gas, or assuming that the
ISM has built up its Fe abundance relatively recently from substellar
values either as the ISM mass itself has increased (for a closed box),
or in concert with strong outflows. However, the ISM abundance pattern
is poorly reproduced in such models for virtually all elements
measured in the NGC 4472 ISM irrespective of the assumed
$[\alpha/Fe]_{\rm stars}$.

While we consider only a single galaxy, large ISM-to-stellar Fe
abundance ratios have not been measured in any elliptical galaxy (see
references in Section 1.2). There is thus a conflict between the
canonical elliptical galaxy SNIa rate \citep{cap97} and the Fe
abundance in the hot ISM of ellipticals. One avenue of reconciliation
is to presume that the {\it effective} rate of SNIa enrichment is much
less than the actual rate because SNIa are not efficiently well-mixed
into the ISM. This implies that there is large amount of missing Fe.

It is plausible that there is substantial galaxy-to-galaxy variation
in the current SNIa rates of elliptical galaxies. Galaxies that are
radio loud or have blue colors have been shown to have enhanced SNIa
rates \citep{man05,del05}. Currently quiescent galaxies may be
sampling the tail-end of a SNIa delay-time distribution with a peak at
one or more timescales much less than the Hubble time, and the rates
may be sensitive to the strength and lookback time of minor star
formation episodes. If this is the case, galactic winds might be
induced in those galaxies with the highest rates, rendering them gas
poor and X-ray dim, and creating an observational bias tilted towards
ellipticals with low SNIa rates.
 
There are robust anomalies in the abundance pattern that cannot be
explained using any of the standard yields (Figures 11 and 12), and
this is the case even if we narrow our consideration to those elements
with spectral features well-resolved by {\it Suzaku}. The ISM
abundances of O, Ne, and Mg are overwhelmingly determined by SNII
ejecta incorporated into, and then ejected, by evolved stars.
Therefore, the NGC 4472 O underabundance - universally replicated in
other elliptical galaxy hot \citep{hb06,ji09} and warm \citep{ab09}
ISM -- can only be explained by decreasing the O abundance in mass
losing stars that, in turn, implies a reduction in (IMF-weighted
average) SNII yield by about a factor of 2. This is not physically
unreasonable considering the sensitivity of SNII production of O to
the treatment of convection in the progenitor stars
\citep{glm97}. Alternatively, one could reduce the predicted O yields
by a truncation of the IMF, since O yields increase steeply with
progenitor mass; however, the mass dependence of Ne and Mg yields are
nearly as steep (e.g., Loewenstein and Mushotzky 1996). These results
for O are especially significant since the [O/H] abundance in HII
regions is often used as a proxy for overall metallicity in studies of
galaxies with star formation \citep{gar02, kk04}. Interpretation of
such data assuming solar ratios among $\alpha$-elements and standard O
yields should be considered with caution, as should calibration of
SN-induced feedback in semi-analytic models that are based on observed
O abundances.

Another evident abundance pattern anomaly manifests as simultaneously
high ISM abundances of Ar, Ca, and Ni. While application of W7 models
can explain the high Ni/Fe ratio, it underpredicts Ar and Ca (Figure
11); the converse is true for models utilizing the WDD1 SNIa
yields. For the $r=1$, $[\alpha/Fe]_{\rm stars}$=0.15 and 0.25 models
with W7 SNIa yields, stellar mass loss accounts for $>80$\% of O, Ne,
Mg, Si, S, Ar, and Ca, $\sim$ 60\% of Fe, and $\sim 50$\% of Ni in the
hot ISM with the remaining metals originating from direct enrichment
via SNIa. To ``fix'' the Ca and Ar discrepancy by increasing the
stellar mass loss abundances through a boost in the SNII Ar and Ca
yields is not very appealing, since Ca is not overabundant (with
respect to Mg) in red-sequence galaxies \citep{gra07,smi09} -- in
fact, Ca traces Fe more closely than it does Mg \citep{gra07}, which
argues for an important SNIa contribution to Ca. In Table 2 and Figure
13, we show the output for an $r=1$ ($\theta_{\rm SNIa}\sim 0.25$),
$[\alpha/Fe]_{\rm stars}$=0.15 model with the following {\it ad hoc}
yield adjustments: a decrease in the average SNII O yield by a factor
2, and SNIa yields for Ca, Ar, Fe, and Ni of 0.05, 0.05, 0.5, and 0.07
$M_{\odot}$, respectively. One can see that, in such models, direct
SNIa enrichment and enrichment via stellar mass return from material
originating in SNIa and in SNII all make significant contributions to
these elements, while the abundances of the lower-Z elements
predominately originate in SNII products incorporated into stars. This
is not meant as a definitive proposal, but to illustrate the sort of
adjustments required. Should such abundance pattern anomalies prove
prevalent, genuine constraints on supernova yields may be derived
(Loewenstein, et al. in preparation).

\begin{deluxetable}{lcccccc}
\tabletypesize{\scriptsize}
\tablewidth{0pt}
\tablecaption{{\it Ad hoc} Model Yields and Origins}
\tablehead{\colhead{} & \colhead{$y_{\rm SNIa}$} & \colhead{$y_{\rm SNII}$} &
\colhead{stars} & \colhead{stars:SNIa} & \colhead{stars:SNII} & \colhead{direct:SNIa}}
\startdata 
O & 0.14  & 0.7   & 0.97 & 0.021 & 0.94 & 0.034 \\
Ne & 0.0045 & 0.38  & 1.0  & 0.0013 & 1.0  & 0.0021 \\
Mg & 0.0086 & 0.12  & 0.99 & 0.0080 &.0.98 & 0.013 \\
Al & 0.00099 & 0.014  & 0.99 & 0.0074 & 0.98 & 0.012 \\
Si & 0.15  & 0.11  & 0.82 & 0.11  & 0.71 & 0.18  \\
S & 0.086  & 0.047  & 0.79 & 0.13  & 0.65 & 0.21  \\
Ar & 0.05  & 0.0077 & 0.60 & 0.25  & 0.35 & 0.40  \\
Ca & 0.05  & 0.0054 & 0.55 & 0.28  & 0.27 & 0.45  \\
Fe & 0.5   & 0.083  & 0.61 & 0.24  & 0.36 & 0.39  \\
Ni & 0.07  & 0.041  & 0.49 & 0.32  & 0.17 & 0.51  \\
\enddata 
\end{deluxetable}

\begin{figure}
\plotone{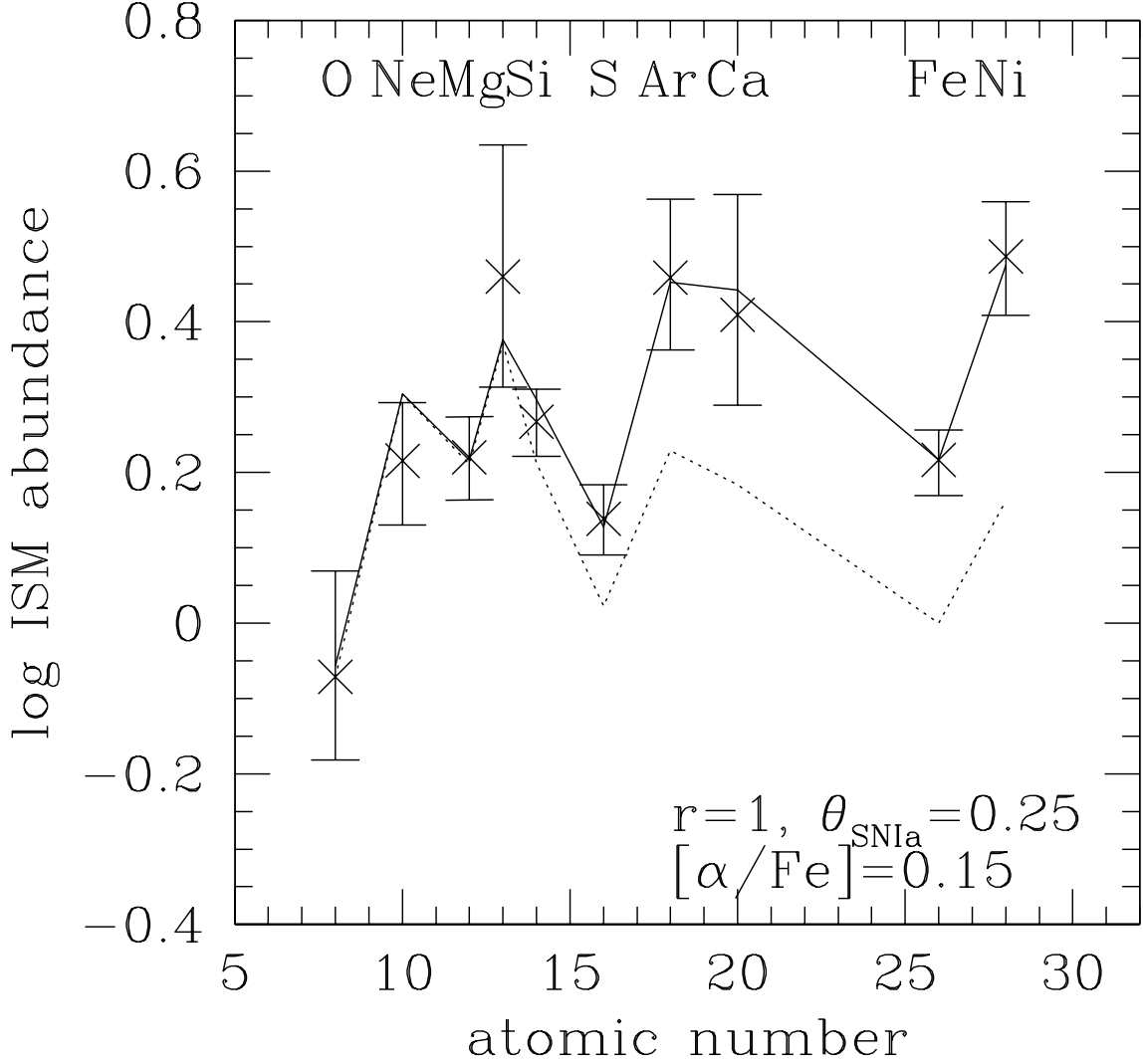}
\caption{Comparison of abundances (including Al -- located between Mg
and S), relative to Fe, in the ISM (crosses with errorbars) and those
predicted in a model where {\it ad hoc} yields for some elements (see
text, Table 2) are adopted for $r=1$, $[\alpha/Fe]_{\rm stars}$=0.15
($\theta_{\rm SNIa}\sim 0.25$). The latter is indicated by the solid
line; stellar abundances are shown by the dotted line.}
\end{figure}

\section{Summary and Conclusions}

We have derived the hot ISM abundance pattern in inner ($0-2.3R_e$)
and outer ($2.3-4.6R_e$) regions of NGC 4472 from analysis of {\it
Suzaku} spectra and used them to study the chemical evolution in this
elliptical galaxy. The low {\it Suzaku} background and relatively
sharp spectral resolution of the {\it Suzaku} XIS detectors enabled us
to extend the range of accurately measured abundances beyond what is
feasible with {\it Chandra} or {\it XMM-Newton} to S, Ar, Ca, and
possibly Al -- while we find general agreement with overlapping {\it
XMM-Newton} results. The abundances of Ne, Mg, Si, S, Fe, and Ni may
be explained by enrichment via a combination of $\alpha$-element
enhanced stellar mass loss and direct injection of SNIa with W7 yields
exploding at a current rate of 0.027 SNU; however, additional SNIa
production of Ca and Ar is required. Ca and Ar are reproduced in
models (with a SNIa rate of 0.035 SNU) where WDD1 yields are adopted,
but at the price of disrupting the previous concordance for S and Ni.

Models with the standard SNIa rate of 0.16 SNU badly overpredict Fe,
unless the enrichment from stellar mass loss is diluted. We introduced
models where this is a consequence of outflow biased in favor of high
metallicity, or an inertial effect resulting from the preexistence
of a metal poor ISM. However, such models predict an underabundance of
elements primarily synthesized in SNII (Ne and Mg for W7 and WDD1 SNIa
yields, and Si, S, Ar, and Ca as well for W7) and offer a poorer match
to the observed abundance pattern.

We confirm previous measurements of a low O abundance; the low O/Mg
and O/Ne ratios imply that the SNII yields in standard models
overproduce O by $\sim 2$.      

Analysis of additional {\it Suzaku} and {\it XMM-Newton} spectra of
elliptical galaxies will be used to investigate the universality of
these results, and probe for diversity in the star formation and SNIa
histories of giant elliptical galaxies. Application of more rigorous
chemical evolution models will sharpen these conclusions, and should
yield new information on the formation and evolution of these systems
and indirect constraints on the physics of Type Ia supernovae.

\acknowledgments

\appendix

\section{Simple Models for Elliptical Galaxy Hot ISM Abundances}

\subsection{Formalism}

In this paper we apply a simplified chemical evolution formalism to
the NGC 4472 observations. A more comprehensive treatment and
application of the coupled chemical evolution of the gas and stars in
elliptical galaxies is in progress.

For a single-phase interstellar medium in a passively evolving galaxy
-- good approximations for giant elliptical galaxies in general, and
NGC 4472 in particular -- the equations for the conservation of
overall ISM mass, $M_{\rm ISM}$, and mass of the $ith$ element,
$f^i_{\rm ISM}M_{\rm ISM}$ where $f^i_{\rm ISM}$ is the mass
fraction, are as follows:

\begin{equation}
{{dM_{\rm ISM}}\over {dt}}=\dot M_{\rm MR}-\dot M_{\rm out},
\end{equation}

and

\begin{equation}
{{d(f^i_{\rm ISM}M_{\rm ISM}})\over {dt}}=G^i_{\rm SNIa}+G^i_{\rm
MR}-(1+b^i_{\rm out})\dot M_{\rm out}f^i_{\rm ISM}.
\end{equation}

The only significant internal source of gas is stellar mass return
(injected at the rate $\dot M_{\rm MR}$), while metals are provided by
Type Ia supernovae (rate $G^i_{\rm SNIa}$) as well as stellar mass
loss (rate $G^i_{\rm MR}$). $\dot M_{\rm out}$ accounts for gas
flowing in or out of the optical galaxy (and is positive for
outflow). The ``bias'' factor $b^i_{\rm out}$ allows for an offset
between the in- or outflowing gas metallicity and the
mass-weighted-average inside the galaxy; henceforth we assume this is
identical for all elements and suppress the index. For $b_{\rm
out}>(<) 0$ winds serve to reduce (increase) the overall galactic ISM
metallicity, while for $b_{\rm out}=0$ winds remove metals but do not
change the overall metallicity. For inflow, $b_{\rm out}>(<)0$
increases (reduces) the galactic ISM metallicity,

These equations may be integrated from the end of the galaxy formation
era $t_{\rm in}$ (at which time it is assumed that the galaxy is fully
assembled and star formation has ceased) to the present day $t_{\rm
 now}$ (13.7 Gyr). It is sensible to ``start the clock'' at $t_{\rm
 in}$ since we know that the ISM mass inside the optical galaxy today
is much less than the stellar mass return rate integrated over the
galaxy lifetime. The star formation era {\it must} have been
accompanied by prodigious outflows in a multiphase ISM, and cannot be
modeled using the formalism developed here. Following \cite{cio91},
and utilizing main sequence turnoff masses and remnant masses from
\cite{sch92} and \cite{sal09}, respectively, we derive $\dot M_{\rm
 MR}\approx AL_B(t/t_{\rm now})^{-\alpha}$, where $\alpha\approx1.3$,
$A\approx 2.4~10^{-11}\theta_{\rm MR}~M_{\odot} {\rm
 yr}^{-1}~L_{B\odot}^{-1}$ and $\theta_{\rm MR}=1$ for a simple
stellar population with present-day blue luminosity $L_B$. In this
simple case $G^i_{\rm MR}=\dot M_{\rm MR}f^i_{\rm stars}$, where
$f^i_{\rm stars}$ is the (constant after $t_{\rm in}$) mass fraction
of the $ith$-element in evolved (mass-losing) stars. The SNIa
enrichment term $G^i_{\rm SNIa}=\dot N_{\rm SNIa}y_{\rm SNIa}^i$,
where $y_{\rm SNIa}^i$ is the $ith$ element SNIa mass yield. The SNIa
rate may be characterized as $\dot N_{\rm SNIa}\approx BL_B(t/t_{\rm
 now})^{-\beta}$ with $B=1.6~10^{-13}\theta_{\rm SNIa}~{\rm
 yr}^{-1}~L_{B\odot}^{-1}$; $\theta_{\rm SNIa}=1$ corresponds to the
rate of \cite{cap97}. \cite{cio91} adopt $\beta=1.1$ and $\theta_{\rm
 SNIa}=1$, while in the models of \cite{rbf09} substantially lower
rates, and a somewhat steeper decline, are derived from binary
population synthesis. In some models we assume that there is an
outflow that is driven by SNIa and given by $\dot M_{\rm out}=K_{\rm
 out}\dot N_{\rm SNIa}M_{\rm ISM}$, where $K_{\rm out}$ is a
dimensionless constant.

\subsection{Special Cases with Exact Solutions}

It proves convenient to introduce the following dimensionless
variables: $\tau\equiv t/t_{\rm now}$, $\mu\equiv M_{\rm ISM}/M_{\rm
ISM}(t_{\rm now})$, $\phi^i=f^i_{\rm ISM}/f^i_{\rm stars}$, $\psi=\dot
Mt_{\rm now}/M_{\rm ISM}(t_{\rm now})$, $\gamma^i=G^it_{\rm
now}/(M_{\rm ISM}(t_{\rm now})f^i_{\rm stars})$. Equations (A1) and
(A2) become

\begin{equation}
{{d\mu_{\rm ISM}}\over {d\tau}}=\psi_{\rm MR}-\psi_{\rm out},
\end{equation}

and

\begin{equation}
{{d(\phi^i_{\rm ISM}\mu_{\rm ISM}})\over {dt}}=\gamma^i_{\rm
SNIa}+\gamma^i_{\rm MR}-(1+b_{\rm out})\psi_{\rm out}\phi^i_{\rm
ISM},
\end{equation}
with $\mu(1)=1$ and the following boundary conditions at $\tau_{\rm
in}=t_{\rm in}/t_{\rm now}$: $\mu_{\rm in}\equiv M_{\rm ISM}(t_{\rm
in})/M_{\rm ISM}(t_{\rm now})$ and $\phi_{\rm ISM,in}^i\equiv f_{\rm
ISM}^i(t_{\rm in})/f^i_{\rm stars}$. The source terms are $\psi_{\rm
MR}=\psi_o\tau^{-\alpha}$, $\gamma^i_{\rm MR}=\psi_{\rm MR}$,
$\gamma^i_{\rm SNIa}=\epsilon^i\eta_o\tau^{-\beta}$, with
$\psi_o=A(M_{\rm ISM}(t_{\rm now})/L_B)t_{\rm now}$, $\eta_o=B(M_{\rm
ISM}(t_{\rm now})/L_B)t_{\rm now}$, and $\epsilon^i=y_{\rm
SNIa}^i/f^i_{\rm stars}$; for SNIa-driven galactic winds $\psi_{\rm
out}=\kappa\eta_o\tau^{-\beta}\mu$, where $\kappa=K_{\rm out}M_{\rm
ISM}(t_{\rm now})$. $Q^i=\eta_o\epsilon^i/\psi_o=(B/A)\epsilon^i$ and
$W=\kappa\eta_o/\psi_o=(B/A)\kappa$ are convenient dimensionless
quantities that factor into determining the importance of SNIa metal
enrichment and outflows, respectively.

There are two classes of simple analytic solutions to equations (A3)
and (A4) -- closed box models where $\dot M_{\rm out}=0$, and
``tracking'' solutions where $\alpha=\beta$. Subsets of the latter
include steady-state solutions (only exactly possible if
$\alpha=\beta$), and solutions where the ISM mass is in a steady
state, but its metallicity is not (i.e., abundances are evolving much
more quickly than the total mass).

For closed box models, the dimensionless abundances at the present day
are given by
\begin{equation}
\phi^i_{\rm ISM}(1)=1-\mu_{\rm in}(1-\phi_{\rm ISM,in}^i)+
{{Q^i\psi_o}\over {(1-\beta)}}(1-\tau_{\rm in}^{1-\beta}),
\end{equation}
where $\mu_{\rm in}$ or $\tau_{\rm in}$ may be eliminated using
\begin{equation}
\mu_{\rm in}=1-{{\psi_o}\over {(1-\alpha)}}(1-\tau_{\rm
in}^{1-\alpha}).
\end{equation}
This takes a particularly simple form for the reasonable initial
conditions $\phi_{\rm ISM,in}^i=1$ (identical ISM and stellar
metallicities) and/or $\mu_{\rm in}=0$ (galaxy evacuated of gas at the
close of the star formation era).

A steady state is possible if $\alpha=\beta$. The dimensionless ISM
mass fraction at all times in this case is given by
\begin{equation}
 \phi^i_{\rm ISM}={{(1+Q^i)}\over {(1+b_{\rm out})}}.
\end{equation}

We also consider the general $\alpha=\beta$ case with SNIa-driven
outflow and $b_{\rm out}=0$. In this case, the solution is
\begin{equation}
\phi^i_{\rm ISM}(1)={{(1+Q^i)(1-\mu_{\rm in})+\mu_{\rm in}\phi_{\rm
ISM,in}^i(1-W)}\over {1-W\mu_{\rm in}}},
\end{equation}
where $W$ and $\mu_{\rm in}$ are related to $\tau_{\rm in}$ via
\begin{equation}
\mu_{\rm in}=W^{-1}\left\{1-(1-W)exp[W\psi_o(1-\tau_{\rm
in}^{1-\alpha})/(1-\alpha)]\right\}
\end{equation}

\subsection{General Features and Application}

All of the simple models above have solutions of the form
\begin{equation}
\phi^i_{\rm ISM}(1)=r^i+sq\epsilon^i, 
i.e.
\end{equation}
\begin{equation}
f^i_{\rm ISM}(1)=r^if^i_{\rm stars}+sqy_{\rm SNIa}^i,
\end{equation}
where $q=B/A$ is proportional to the present-day ratio of SNIa to mass
return rate. For the simplest steady state models ($r^i=s=1$) mass
return enriches the ISM to the stellar metallicity (first term), and
SNIa further enhances the abundances according to the ratio of
supernova to mass-loss rates and the SNIa yield. In the other models
considered above $r$ represents dilution or amplification of stellar
mass return due to initial conditions for closed box models, biased
outflow in the steady-state $b_{\rm out}\ne 0$ case, and the interplay
between initial conditions and outflow for the tracking solutions. The
factor $sq$ represents the SNIa enrichment and depends on the ratio of
supernova to mass-loss rates with these same causes of
dilution/amplification for the steady-state $b_{\rm out}\ne 0$ and
tracking solutions, and modification due to the mismatch in the SNIa
and stellar mass loss time dependencies for the close box case.

In particular (see Table 3) $r^i=(1+b_{\rm out})^{-1}$ for steady
state, $r^i=1-\mu_{\rm in}(1-\phi_{\rm ISM,in}^i)$ for closed box, and
$r^i=[(1-\mu_{\rm in})+\mu_{\rm in}\phi_{\rm
ISM,in}^i(1-W)]/(1-W\mu_{\rm in})$ for tracking solutions. The SNIa
parameter $s=(1+b_{\rm out})^{-1}$ for steady state,
$s=\psi_o(1-\tau_{\rm in}^{1-\beta})/(1-\beta)$ for closed box, and
$s=(1-\mu_{\rm in})/(1-W\mu_{\rm in})$ for tracking solutions.

\begin{deluxetable}{lccc}
\tabletypesize{\scriptsize}
\tablewidth{0pt}
\tablecaption{Definitions for $r$ and $s$ in Different Models}
\tablehead{\colhead{Model} & \colhead{$r^i$} & \colhead{$s$}}
\startdata 
Steady state & $(1+b_{\rm out})^{-1}$ & $(1+b_{\rm out})^{-1}$ \\
Closed Box & $1-\mu_{\rm in}(1-\phi_{\rm ISM,in}^i)$
& $\psi_o(1-\tau_{\rm in}^{1-\beta})/(1-\beta)$\\
Tracking & $[(1-\mu_{\rm in})+\mu_{\rm in}\phi_{\rm
ISM,in}^i(1-W)]/(1-W\mu_{\rm in})$ & $(1-\mu_{\rm in})/(1-W\mu_{\rm
in})$\\
\enddata 
\end{deluxetable}

One can relate the ISM abundances of the $ith$ and $jth$ elements,
eliminating the factor $s$, via

\begin{equation}
f^i_{\rm ISM}(1)=rf^i_{\rm stars}+(f^j_{\rm ISM}(1)-rf^j_{\rm
stars}) {{y_{\rm SNIa}^i}\over {y_{\rm SNIa}^j}}
\end{equation}
or
\begin{equation}
{{f^i_{\rm ISM}(1)}\over {f^j_{\rm ISM}(1)}}=
{{f^i_{\rm stars}}\over {f^j_{\rm stars}}}
{{f^j_{\rm stars}}\over {f^j_{\rm ISM}(1)}}r+
\left(1-r{{f^j_{\rm stars}}\over {f^j_{\rm ISM}(1)}}\right)
{{y_{\rm SNIa}^i}\over {y_{\rm SNIa}^j}},
\end{equation}
where we have made the simplifying assumption that $r^i=r$ is the same
for all elements (already implied for steady state solutions by
adopting constant $b_{\rm out}$, and otherwise equivalent to constant
$\phi_{\rm ISM,in}^i$ -- i.e., an element-independent initial ratio of
ISM-to-stellar abundances).

In the approach we adopt here, we calculate the ISM abundance pattern
for a given stellar abundance pattern (corresponding to a particular
$[\alpha/Fe]$ ratio; see below) and ISM-to-stellar Fe abundance, for a
range of values of $r$. One then compares the pattern with the NGC
4472 observations to see if consistent scenarios emerge and to
identify any anomalies. The parameter $r$ may then be interpreted {\it
 post facto} under the different classes of models, as can $s$ for
values of the stellar and ISM Fe abundances (and not just the
ratios). We consider the SNIa and SNII+HN yields from
\cite{nom97,kob06}, and use these to set the stellar abundance pattern
for a given value of $[\alpha/Fe]_{\rm stars}$, where $[\alpha/Fe]$ is
defined as follows. We calculate the ratio, with respect to the Fe
mass fraction, of the sum of the mass fractions of all elements under
consideration lighter than Fe, divide by the same ratio calculated
using the \cite{gs98} solar standard, and take the logarithm. For a
given set of yields, $[\alpha/Fe]$ corresponds to a particular number
ratio of SNIa to SNII that contributed to establishing the stellar
abundances. The \cite{kob06} SNII+HN yields are averaged over a
Salpeter IMF with maximum mass 50 $M_{\odot}$, and the hypernova
fraction set to match the abundances of low metallicity stars in the
Milky Way halo. Because of the rapid buildup of metals in elliptical
galaxies, we adopt the solar metallicity SNII+HN yields of
\cite{kob06}. Selection of a different metallicity would have little
effect on the final stellar abundance pattern and corresponding value
of $[\alpha/Fe]$ (compared to the uncertainty in the yields
themselves; Gibson et al. 1997), and primarily alter the corresponding
relative numbers of SNII and SNIa. The effects of different yield sets
and IMFs is beyond the scope of the present work, but something we
will consider in future investigations. The same set of yields are
used for SNIa that directly enrich the ISM, and those that enrich the
stars.

As it turns out, numerical experiments demonstrate that $s\approx r$
for most reasonable models. This is exactly true for the steady state
models. For closed box models with $\phi_{\rm ISM,in}^i=0$, it
deviates by $<20$\% for $\theta_{\rm MR}=1$, $M_{\rm ISM}(t_{\rm
now})/L_B\sim 0.1$, and $0<\beta<2$. The dependence on $\phi_{\rm
ISM,in}^i$ is flat until one gets to the regime where the solution
corresponds to the ISM Fe abundance being substellar until very
recently and then suddenly being enriched to the observed superstellar
value by a concentrated period of SNIa enrichment. Similarly, for
tracking solutions $s$ differs significantly from $r$ only in cases
where, at a recent epoch the galaxy was filled with a large mass of
low-abundance gas. In this scenario, the galaxy must be experiencing a
powerful outflow that has ejected most of the low metallicity gas and
replenished it with SNIa-enhanced stellar mass return. For example,
using our standard parameters, to obtain $s=0.5$ for $r=1$, one
requires the galaxy have 6.4 times its present ISM mass at $\tau_{\rm
in}=0.7$, and with $<0.08$ solar Fe abundance. {\it This implies that,
for relatively quiescent elliptical galaxies, for a given value of the
dilution factor $r$ the offset between the stellar and ISM Fe
abundances (and the SNIa Fe yield) can be robustly used to estimate
the SNIa to mass return rate. If the SNIa and mass loss rates have
different time-dependencies the ratio will correspond to a ratio of
time-averaged rates, but for standard parameters this does not
drastically differ from the ratio at the present day.}


\clearpage

\end{document}